\documentclass[preprint,journal,hideappendix]{vgtc}            

\onlineid{1758}

\vgtccategory{Research}

\vgtcpapertype{theory/model}

\title{ Character-Oriented Design for Visual Data Storytelling\vspace{-5pt}}

\author{
    \authororcid{Keshav Dasu}{0000-0002-7689-4368},
    \authororcid{Yun-Hsin Kuo}{0009-0000-1891-8993}, and
    \authororcid{Kwan-Liu Ma}{0000-0001-8086-0366}
}

\authorfooter{
  \item Keshav Dasu, Yun-Hsin Kuo, and Kwan-Liu Ma are with the University of California, Davis. E-mail: 
\{kdasu, yskuo, klma\}@ucdavis.edu

}

\abstract{
When telling a data story, an author has an intention they seek to convey to an audience. This intention can be of many forms such as to persuade, to educate, to inform, or even to entertain. In addition to expressing their intention, the story plot must balance being consumable and enjoyable while preserving scientific integrity. In data stories, numerous methods have been identified for constructing and presenting a plot. However, there is an opportunity to expand how we think and create the visual elements that present the story. Stories are brought to life by characters; often they are what make a story captivating, enjoyable, memorable, and facilitate following the plot \revise{until} the end. Through the analysis of 160 existing data stories, we systematically investigate and identify distinguishable features of characters in data stories, and we illustrate how they feed into the broader concept of “character-oriented design”. We identify the roles and visual representations data characters assume as well as the types of relationships these roles have with one another. We identify characteristics of antagonists~\revise{ as well as define conflict} in data stories. We find the need for an identifiable central character that the audience latches on to in order to follow the narrative and identify their visual representations. We then illustrate “character-oriented design” by showing how to develop data characters with common data story plots. With this work, we present a framework for data characters derived from our analysis; we then offer our extension to the data storytelling process using character-oriented design. To access our supplemental materials please visit \url{https://chaorientdesignds.github.io/}.

} 

\keywords{Storytelling, Explanatory, Narrative visualization, Visual metaphor }

\graphicspath{{figs/}{figures/}{pictures/}{images/}{./}} 

\usepackage{tabu}                      
\usepackage{booktabs}                  
\usepackage{lipsum}                    
\usepackage{mwe}                       

\usepackage{mathptmx}                  

\usepackage{microtype}                 
\PassOptionsToPackage{warn}{textcomp}  
\usepackage{textcomp}                  
\usepackage{times}                     
\usepackage{cite}                      
\usepackage{tabu}                      
\usepackage{tabularx}

\usepackage{amsmath,amssymb}
\usepackage{color}
\usepackage{colortbl}
\usepackage[normalem]{ulem}
\usepackage{array}
\usepackage{enumitem}

\usepackage[font={small,sf}]{caption}
\usepackage[font={scriptsize,sf}]{caption}

\begin{document}


\newcommand{\revise}[1]{{\color{black}#1}}
\newcommand{\nosense}[1]{{\color{pink}#1}}
\newcommand{\yunhsin}[1]{{\color{teal}#1}}
\newcommand*{\tabIndent}{\hspace*{0.5cm}}%

\newcommand{\MC}[1]{{\color{MC} \textbf{#1}}}
\newcommand{\SC}[1]{{\color{SC} \textbf{#1}}}
\newcommand{\AC}[1]{{\color{AC} \textbf{#1}}}


\definecolor{SC}{HTML}{2E7D32}
\definecolor{AC}{HTML}{b71c1c}
\definecolor{MC}{HTML}{BF9000}

\teaser{
  \centering
  \includegraphics[width=\linewidth]{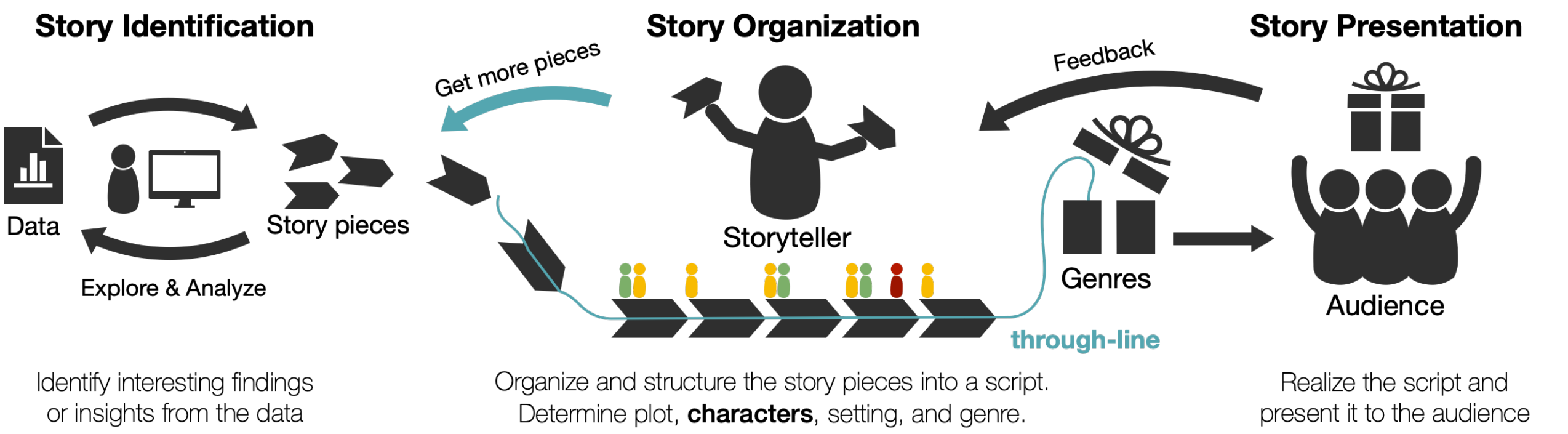}
  \caption{%
  	Character-oriented visual storytelling consists of three stages: story identification, organization, and presentation. A key task in story organization is to determine what \MC{main}, \SC{supporting},  \AC{antagonist} characters, etc. are and their relations to the plot.
  }
  \label{fig:ds_process}
}

\firstsection{Introduction}

\maketitle
Information, at times, can be abstract and intangible, which may lead to difficulties in communication. The beauty of visualization is captured in its ability to make the intangible tangible, the invisible visible, and the inaccessible accessible. Through visualization, we can utilize visual representations to embody complex and often large datasets, reveal hidden insights about both known and unknown phenomena, and afford a means to showcase findings as well as share insights with broader audiences. We, as data storytellers, are concerned with presenting these findings to large audiences. Stories and visual storytelling have been shared and consumed by our earliest ancestors. Some of the earliest forms of visual storytelling~\cite{fritz2007hidden} played a role in communicating where rich sources of food can be located or where to avoid dangerous beasts. In visualization, we have utilized storytelling for a variety of communicative needs since it is effective for engagement~\cite{kang2020role}, memorability~\cite{sarica2016effect,dudukovic2004telling}, and showing casualty~\cite{forster2010aspects}. 

As data storytellers, we play a role in capturing and sharing the \revise{wonder} we see in data with others. In our stories, we are challenged to emphasize the scientific \revise{insights} of our content and simultaneously engross~\cite{isenberg2018immersive} the audience with our narrative. The challenge of ensuring our content is both consumable and enjoyable while preserving scientific integrity constrains our story design. These constraints \revise{can} result in the audience having a difficult time understanding~\cite{dasu2020sea,boy2015storytelling} \revise{insights}, \revise{topic relevancy}, or where in the story to focus. 


In data stories, numerous methods have been identified for constructing and presenting a plot. A story plot~\cite{forster2010aspects} is a narrative of events, with the emphasis falling on causality. The data storytelling process~\cite{lee2015more} can be viewed as three stages \revise{---} identification, organization, and presentation. Typically, the first step resolves in the accumulation of a set of events (i.e., ``story pieces''). These pieces are often the insights derived from either the collaborative efforts of data analysts and domain experts or the automation leveraged by statistics~\cite{srinivasan2018augmenting, wangDataShot2020}. The collection of events is guided by the shared intent of the author and analysts, which is the intention they seek to convey to the audience. This intention can take on many forms~\cite{ojo2018patterns} (e.g., to inform, to educate, to entertain, or to explore) and centers the story. Next, in the organization stage, several narrative frameworks~\cite{lee2015more,segel2010narrative,stolper2016emerging,yang2022} can assist us in sequencing these events into a cohesive story plot. During this stage, we need to ascertain several properties about these events, namely their relationship to one another and their ordering. We should end up having a structured outline of what we want to convey and the sequence in which to present them. Lastly, we have the presentation stage, where we give the look and feel to the story. There are many methodologies~\cite{hullman2013deeper,hullman2011visualization,heer2007animated} at our disposal for tailoring our story for the target audience. 
However, it is within the presentation stage that there is an opportunity to expand how we view and design the visual elements that act out our story plots.

In our work, we are interested in data-driven, visual storytelling, particularly the characters that bring them to life. Data storytellers want to create rich experiences that evoke an emotional response, draw the audience in, and leave them with something to remember. Stories \revise{can be} brought to life by characters; often they make a story captivating, enjoyable, memorable, and facilitate following the plot \revise{until} the end. 
\revise{
In other media, characters are often used as a bridge for the audience to cross into an unfamiliar and perhaps complex new worlds~\cite{truby2008anatomy,campbell2008hero,forster2010aspects}.
For example, in \textit{Star Wars: Episode IV -- A New Hope}, Luke Skywalker, the protagonist, is the bridge that leads the audience into the \textit{Star Wars} world.
Throughout the story, the audience learns more about the setting and the rules of this world (e.g., force) through his behaviors, rather than from a list of terminology.
Through the lens of characters, the audience can gain an understanding of a world without prior knowledge. 
We are inspired to investigate the possibility of applying characters to convey scientific insights in data stories.
A deeper understanding of data characters could address open data storytelling opportunities~\cite{isenberg2018immersive}. 
}

This work seeks to address the following \textemdash{}
what a data character is, \revise{how we classify characters in data storytelling, and a space for how we can develop a data character and apply it within a story.} 
Through the analysis of 160 existing data stories, we \revise{present a framework for data characters, where we identify three fundamental character roles: \MC{main}, \SC{supporting}, and \AC{antagonist} characters.}
\revise{
With this character-oriented design space, we further investigate how ``conflicts'' are contextualized in data storytelling, as they drive the narrative and can elevate a telling of a story~\cite{truby2008anatomy,forster2010aspects}.
Notably, we find that designing an identifiable central character can support the author in aligning the story pieces with their intentions, arranging the sequence with a consistent message, and delivering this message to the target audience.
}

We consider our primary contributions are:
\begin{itemize} [noitemsep,topsep=2pt,parsep=0pt,partopsep=0pt]
    \item a \revise{framework}~\footnote{https://chaorientdesignds.github.io/} for data characters that extend to the data storytelling process;
    \item a summary of storytelling terminology derived from a variety of storytelling and visualization literature as well as an assessment of data characters in the current literature; and 
    \item case studies in various data story genres to demonstrate the applicability of our design space. 
\end{itemize}

\section{Background}
\revise{In this section, we walk through} key storytelling terms and contextualize them for data storytelling.
Storytelling structure and paradigms~\cite{forster2010aspects,truby2008anatomy,campbell2008hero} have been ever present and are constantly evolving with new mediums and formats. However, we are interested in data stories and determining what applies to storytelling in general and how it can be translated and used for our purposes as \textit{data} storytellers. Data stories can be presented in a variety of ways, such as infographics, comics, videos, virtual experiences, and so on.

\subsection{What is a Story?}
People have shared and told stories for ages~\cite{fritz2007hidden} and consequently have postulated the rules and structures for effective storytelling. When considering where to focus on influences for data storytelling\revise{,} we draw on elements from both written~\cite{forster2010aspects,truby2008anatomy} and visual media~\cite{field2005screenplay}. In \revise{\textit{Aspects of the Novel} by E. M. Forster}~\cite{forster2010aspects}, he analyzed the common aspects that all English-language novels share: \textit{story}, \textit{people}, \textit{plot}, \textit{fantasy}, \textit{prophecy}, \textit{pattern}, and \textit{rhythm}. He describes a \textbf{story} as ``a narrative of events arranged in their time sequence'' and the \textbf{plot} as “also a narrative of events, with the emphasis falling on causality”. Naturally, we look to see how these principles translate into data stories. With written and visual media, authors typically have more freedom and flexibility when creating their stories.

Data stories, however, tend to have less flexibility in that they often are constrained by ``non-alterable non-fiction''~\cite{isenberg2018immersive}. Still, from existing literature~\cite{isenberg2018immersive,bran2010message}, we can surmise that common themes for both storytelling and data storytelling are: \textit{characters}, \textit{plot}, \textit{theme}, \textit{setting}, and \textit{conflict}. The goal of most \textbf{data stories} is to reach a wide or targeted audience through the presentation of visualized findings or messages. The role of a storyteller~\cite{truby2008anatomy} is to summarize all the narrative events such that the audience perceives this as a self-contained story.

\begin{table*}[t]
\small
\centering
\begin{tabular}{p{0.01\textwidth}>{\raggedright}p{0.175\textwidth}>{\raggedright}p{0.33\textwidth}p{0.37\textwidth}}
    \toprule
    \multicolumn{2}{p{1.5625in}}
    {\textbf{Storytelling Terminology}} & 
    \textbf{Description} &
    \textbf{Contextualized for Data Storytelling}  \\
    \arrayrulecolor{black!30} \midrule
    
    
    \multicolumn{2}{p{1.5625in}}{\textbf{Story} or Narrative} &
    Consists of many subsystems working together (e.g, characters, plot, and theme). It is an account of events arranged in their time sequence~\cite{forster2010aspects}. &
    A combination of visualized findings or messages with connections such as temporal or causal relations~\cite{isenberg2018immersive,gershon2001storytelling}.\\
    
    &
    \textbf{Theme} &
    A recurring idea. A story can have many themes. &
    These appear as concepts paired with an intention. (i.e., to inform, persuade, entertain, comfort, explain, or terrorise~\cite{ojo2018patterns}).\\
    
    &
    \tabIndent Through-Line &
    A single theme that runs from the start to the end of a story. It interweaves the roles of the characters with the plot. &
    One primary concept and intention that drives the story and motivates the characters' actions~\cite{ojo2018patterns}. \\
    
    
    &
    \textbf{Plot} &
    A combination of events and how those events are revealed. The arrangement (or the sequence) of events in the story.  &
    A \revise{causal} relationship~\cite{lee2015more} between a set of events, depicted by the \revise{behaviors} and interactions~\cite{ma2012scientific} of data-driven visual elements.
    \\
    
    &
    \tabIndent Event or Story Piece & 
    Atomic element of a story. It may focus on one character's status or \revise{behavior}, or the relationships among multiple characters.  &
    ``Story pieces''~\cite{lee2015more} that are derived from data, provided by data analysts and domain experts\revise{, e.g., data facts and human insights}.
    \\
    
    &
    Genre &
    Classification and organization of works into categories. &
    Magazine, annotated chart~\cite{hullman2013contextifier,bryan2016temporal,ren2017chartaccent}, partitioned poster, flow chart, comic strip~\cite{bach2018design}, slide show~\cite{wangNarvis2019}, and film/video/animation~\cite{yang2022,amini2015understanding,bradbury2020documentary} \\
    
    &
    Setting &
    A cluster of actual states of affairs or various events where the story takes place. & 
    The devices and location where the story is presented~\cite{lee2015more} and the various scientific domains~\cite{ma2012scientific} where the insights were derived from.\\

    \multicolumn{2}{p{1.5625in}}{Storytelling} &
    To both share and provide an experience to an audience. To give the audience a form of knowledge~\cite{truby2008anatomy,field2005screenplay} that is both emotional and entertaining. &
    \revise{This is the same, with an additional condition that the story must be based on the data.} \\
    
    \multicolumn{2}{p{1.5625in}}{Audience} &
    The targeted group of people who will receive the story &
    This is the same in data storytelling. \\

   \bottomrule

\end{tabular}
\caption{
Terminology with the definitions contextualized for data storytelling. These terms and their mappings were derived from a breadth of visual and written literature~\cite{truby2008anatomy,field2005screenplay,fink2014dramatic,forster2010aspects}, corroborated with the narrative and data storytelling literature~\cite{segel2010narrative,ma2012scientific,shi2018meetingvis,lee2015more,isenberg2018immersive,ojo2018patterns}.
} 
\label{table:storytelling_defs}
\end{table*}

\subsection{Elements of a Story}
The terminology for data storytelling as it relates to storytelling has some differences. We define a set of storytelling terms for consistency and clarity in~\revise{\autoref{table:storytelling_defs}}. However, even within data stories, there are many varying explanations~\cite{ma2012scientific,shi2018meetingvis,lee2015more,bran2010message,isenberg2018immersive} of what \textit{through-line, plot}, and \textit{character} mean. Therefore, we provide a deeper explanation of those in this subsection. 

\vspace{2pt}
\noindent\textbf{Through-Line.} When we look at the storytelling process~\cite{truby2008anatomy,pixar22coats,field2005screenplay} and data storytelling process~\cite{lee2015more}, both require ascertaining what the author wants to offer the audience. 
As authors of data stories, we want the audience to feel something~\cite{isenberg2018immersive} as they consume our story and leave with something~\cite{ojo2018patterns} when they are finished. 
\revise{It is important for us to identify the core message that we want the audience to come away with.
\revise{For example}, a data story could seek to persuade the audience to take action against climate change by communicating the impact microplastics have on biodiversity. 
Alternatively, the story could intend to explain the flaws of a misconception, all types of plastic have the same effect on the environment. In both scenarios, the authors may communicate the life cycle of microplastics to the audience, while their intention influences the narrative.
}
This \revise{message} is often referred to as the theme of a story, or a recurring idea. A story can have many themes \revise{(e.g., biodiversity)} but a through-line~\cite{field2005screenplay} is the \textit{theme} \revise{(e.g., the impact of microplastics)} that runs from the start to the end of a story. It is what helps keep the story on track and helps drive it from start to finish. A through-line interweaves the roles of the characters with the plot. It helps ensure consistency and continuity for the entire story. 

\vspace{2pt}
\noindent\textbf{Plot.} A plot is a description of a set of events with a purpose~\cite{truby2008anatomy}. Each event that transpires is causally connected and each event is essential to the overall story. As Poe~\cite{shen2008edgar} finds with his theory of ``unity of effect'', every element of a story should help create a single emotional impact. Namely, every element in our story must tie in with the through-line.

\vspace{2pt}
\noindent\textbf{Character.} A character is an entity that influences \revise{either itself} or others. A story typically tracks what an entity wants~\cite{campbell2008hero}, what the 
entity will do to get it, and what cost the entity will have to pay along the way. What pulls the character along is ``desire''. A character will take action in pursuit of a desire while learning new information about this desire. This new information causes a change in the course of actions and hence influences the story. A character in pursuit of a desire always encounters a struggle, which may cause a change in the character itself. The task of the storyteller is to present a change in a character or illustrate why that change did not occur. The task of this work is to show how this same process can be applied to data stories.

\section{Data Storytelling}
Within visualization, there is a large body of work~\cite{lee2015more,tong2018storytelling} that utilizes and documents storytelling with data. Data storytelling and narrative visualization are two notable branches. Recently, the literature contained in these branches was further organized into three groups~\cite{tong2018storytelling}. The works are sorted based on whether they (1) \textit{address who are the main subjects involved} (authoring tools and audience), (2) \textit{assess how stories are told} (narratives and transitions), or (3) \textit{consider why storytelling is effective for visualization} (memorability and interpretation). This work seeks to contribute to \textit{how stories are told} by systematically investigating and identifying distinguishable features of characters in data stories. The general storytelling process~\cite{truby2008anatomy,field2005screenplay} considers both the plot and the characters. We find little has been done to address the design of data characters\revise{, as }there is a current focus on creating and conveying the data story plot. In this section, we review the notion of a data character and how it has been previously considered.

\subsection{Data Storytelling Process}
\label{sec:datastoryprocess}
To better see the relationship between the storyteller and the data storyteller, we organize data storytelling literature into three stages, as shown in~\autoref{fig:ds_process}, based on the process \revise{of the data storyteller}~\cite{lee2015more} \revise{---} identification, organization, and presentation of the story. 
At the beginning of developing a story, the data storyteller may not know what to discuss or share yet. 
\revise{A set of events (or story pieces~\cite{lee2015more}), such as data facts or human insights, should be identified beforehand so that the data storyteller can figure out what to share.}

\revise{Merely having a set of story pieces to report does not give a story. Through the relationships among story pieces (e.g., causality), the data storyteller can organize the story pieces into a story plot for presentation.
However, }if the storyteller is unclear on the relationships or why the subject matter is worth presenting, the audience \revise{may} feel similarly when consuming the story. \revise{The organization stage is where data storytellers select story pieces, sequence them into a story plot, and leave with a structured outline for the story.}
Several frameworks offer systematic ways to sequence a story plot for various genres~\cite{ojo2018patterns,segel2010narrative,yang2022,bach2018design, hullman2011visualization}. Here\revise{,} we review the \revise{relevant} literature on story plots from two aspects: the \revise{narrative} structure and the communicative goal.

There is a \revise{popular} set of three narrative structures~\cite{segel2010narrative} for data storytelling\revise{:} the drill-down, the martini glass, and the interactive slideshow, which have been \revise{heavily discussed in this space.}
Recently, Yang et al.~\cite{yang2022} \revise{provide} guidelines for applying Freytag's Pyramid, a narrative structure that has been widely used in film and literature, to data stories.
These structures are similar to what Truby describes as the \revise{\textit{story movement}} in the space of written storytelling~\cite{truby2008anatomy}.
\revise{Truby} describes a set of \revise{narrative structure} patterns \revise{that} storytellers draw from nature to connect elements in a sequence (e.g., linear, meanderings, spiral, and branching).
For data stories, \revise{the} narrative structure addresses \revise{the} story movement, \revise{yet the pace of this movement is often influenced by the audience's interactions~\cite{segel2010narrative}.}
\revise{For example, t}he drill-down structure can present multiple \revise{plot branches} and even different stories \revise{together}, thus enabling the audience to explore the story.
\revise{Meanwhile, }the alternatives often present \revise{a single linear narrative}, which is easier for the author to maintain the plot consistency, and therefore has been the focus of automated reasoning~\cite{hullman2013deeper, kim2017graphscape}.
\revise{In terms of the communicative goal, }Adegboyega and Heravi~\cite{ojo2018patterns} identify seven types of story plots from data stories. Their work provides a great jumping-off point for \revise{story} creation.
\revise{In~\autoref{sec:relationships}, we expand on their work and} describe generically how \revise{data} characters \revise{may} weave \revise{within these seven story plots.} 


\revise{It is important to note that the genre~\cite{segel2010narrative} and the setting~\cite{lee2015more} of a story should also be settled at this organization stage.}
\revise{Determining a genre (or genres) of the story can help data storytellers narrow down the choices of the ideal story plot and narrative structure, aligning with their intentions better.}
Segel and Heer~\cite{segel2010narrative} identify a set of genres \revise{for data stories}: magazine, annotated chart~\cite{hullman2013contextifier,bryan2016temporal,ren2017chartaccent}, partitioned poster, flow chart, comic strip~\cite{bach2018design}, slide show~\cite{wangNarvis2019}, and film/video/animation~\cite{yang2022,amini2015understanding,bradbury2020documentary,amini2016authoring}.
Akin to other storytelling fields, these genres can be paired with each other.
Many of these genres have been further developed, and methodologies and frameworks for using them have been offered. 
Refer to Tong et al.~\cite{tong2018storytelling} for more details\revise{.}

\revise{During} the organization stage, the data storytellers may begin assessing which visual elements will best achieve their communicative goals and support \revise{advancing} the story.
\revise{Understanding the relevancy of each visual element to the story theme often can help the data storytellers locate important story pieces~\cite{hullman2011visualization}.}
\revise{Irrelevant story pieces in a story can become multiple story sub-plots. Without proper curation, the story likely encounters issues at the presentation stage, such as the audience struggling with comprehending the story and losing interest~\cite{boy2015storytelling, dasu2020sea}.}
\revise{This motivates our work to investigate how visual elements can be modeled as data characters. The characterization of story pieces with data characters may help reinforce the through-line and construct the story plot, supporting the data storytellers in delivering a clear message to the audience with their stories.}

\begin{figure*}[h]
    \centering
    \includegraphics[width=\linewidth]{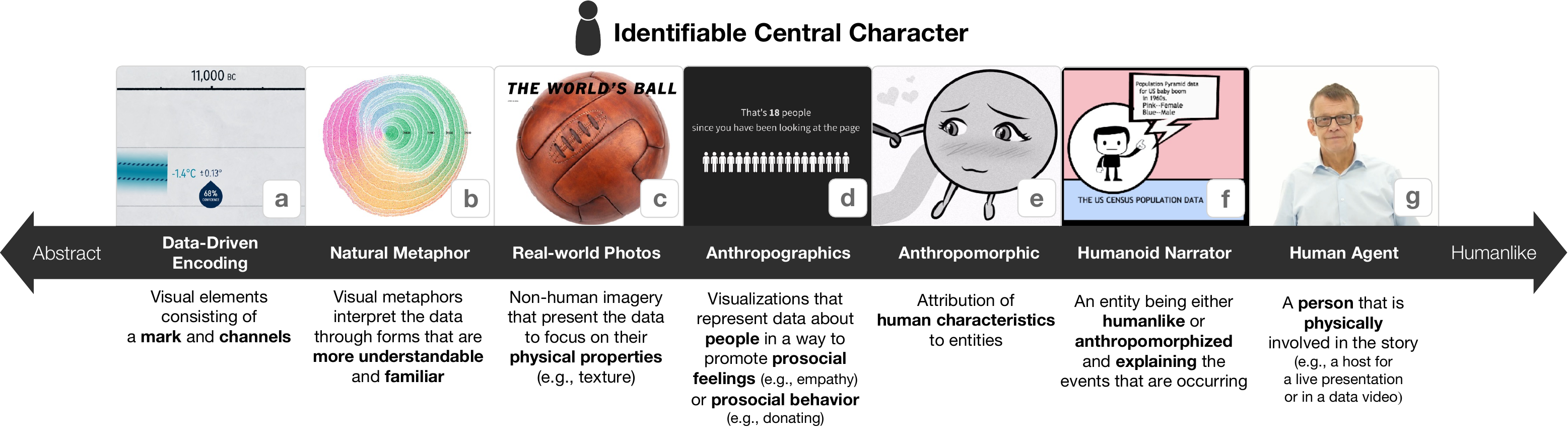}
    \caption{Identifiable Central Character (ICC). In character-oriented visual data storytelling, we require a character that is central to the story and can be visually identified by the audience. From our analysis, we found this character's visual depiction can range from an abstract representation to a person. (a) a visual encoding such as a line to present the temperature~\cite{howsure_2021} (b) Tree rings~\cite{cruz2018process} showing the immigration (c) Soccer ball photos to show the texture progression in panels~\cite{soccer_2014} (d) the anthropographic people~\cite{scarr_2020} denote the loss to COVID-19 (e) a character that represents productivity from a data comic~\cite{WillikinWoolf} (f) a humanlike character explaining the content~\cite{zhao2015data} (g) Professor Hans Rosling~\cite{gapminder} starring in a data video.}
    \label{fig:icc}
\end{figure*}

\subsection{Characters in Data Stories}
Prior research has addressed character-based~\cite{cavazza2001narrative,cavazza2001characters} or character-driven storytelling~\cite{caiCharacterPlot,charles2001character}. 
Cai et al.~\cite{caiCharacterPlot} address the importance of \revise{character-} and plot-driven storytelling and offer a hybrid system involving \revise{both approaches}. 
Cavazza et al.~\cite{charles2001character,cavazza2001narrative,cavazza2001characters} \revise{focus} on character-driven stories and \revise{propose} an engine that models character behaviors. 
Their work \revise{has} a clear focus on automatic narrative generation, \revise{where} characters in all of these works are presented as either human or virtual agents. 
These works focus on improving stories that either are rooted in traditional storytelling (\revise{i.e.,} not data-driven and allow for more creative freedoms of the author) or express the story content using virtual humanoid agents. 
However, data stories often are not expressed using human or virtual agents but rather through abstract representations. \revise{Data} stories are also often rooted in non-alterable non-fiction, constraining the storytelling process. Our work contends that we should view and think of abstract \revise{visual} representations as \revise{data} characters\revise{,} similar to how prior research applies techniques to virtual agents. 

To provide recent examples of what we consider an effective data character, we look to the popular storyline visualization~\cite{storyline2012,gove2021automatic,baumgartl2021}. This visualization depicts an abstract \revise{line} encoding that represents the progress of an entity during a temporal period. We can view this encoding as a character\revise{, drawing the attention of an audience}. Furthermore, it can be used as a through-line and an entry point into understanding the change and growth of a single data-driven entity. Storyline visualizations contain multiple storylines, thus multiple characters, illustrating their relationships. 
Other examples include visualizations with properties that make them more identifiable or personable, such as anthropographics~\cite{morais2020showing,boy2017showing}. 
These representations were shown to be effective in creating an emotional connection with underlying content\revise{~\cite{NegativeEmo2022,boy2017showing}, while other evidence revealed their limitations of eliciting a specific emotion from an audience~\cite{liem2020structure}.}
\revise{
These works motivate a need for the data storytelling community to translate devices and structures from visual storytelling (i.e., film literature).
We feel data characters may serve as a stepping stone to developing and eliciting an emotional connection with the audience, yet to be investigated in future work.
}
A data character can be the visual elements that are ``performing'' the data story~\cite{ma2012scientific}. There are some direct references~\cite{ma2012scientific,shi2018meetingvis,zhao2015data} on the role a data character assumes; however, a story can not exist without characters. 
When \revise{designing} data characters, we are looking for visual entities \revise{that relate to the theme of the story and advance the story plot}. A data character could naturally take on rules of visual encoding, the well-established combination of marks and channels with their respective mappings to data. \revise{A data character should consider properties} that delve deeper into the communicative effort and that help the audience bridge the gap between science and the story. Thus, a foundation for the properties of a data character can be derived from works that address visualization design~\cite{shi2018meetingvis,morais2020showing,boy2017showing,stolper2016emerging,zhao2015data} and visual metaphors~\cite{risch2008role,petridis2019human,li2015metaphoric,cai2015applying,sallaberry2016contact,wang2015design,cruz2018process,dasu2018organic}. There is limited research that addresses the role of a data character in data storytelling. \revise{In the following section, we provide a framework to classify data characters and their behaviors.}


\section{\revise{Data Character Framework}}
\revise{
In this section, we describe the development of our framework for data characters that was derived through discussions with storytelling experts and from the analysis of existing data stories. 
In \autoref{sec:roles}, we present our framework for how we classify data characters and the behaviors we identified.
We then describe a space for how we can develop a data character and apply it within a story.   
We refer to this space as \textit{character-oriented design}, to be elaborated in~\autoref{sec:designspace} and~\autoref{sec:relationships}.
}

\revise{\subsection{Framework Derivation}} 
To better understand the nature of data characters, the forms that they assume, and their behavior in data storytelling, \revise{we created a corpus of 160 data stories and analyzed these stories using a codebook we developed.}  
The corpus was made by merging several other corpora~\cite{hohman2020communicating,NegativeEmo2022,VNF2017,9552203} and updated to include newer stories. The merged corpus size was reduced based on repeats and any stories that were no longer accessible (i.e., required \revise{Adobe Flash Player} or \revise{went offline}). We further filtered down the set to primarily visual stories.
\revise{In these stories the content and core messages are communicated via data-driven visualizations, animations, or videos. We excluded those heavily dependent on} multiple paragraphs of text, where the visualizations served as annotations or figures \revise{rather than} the driving narrative force. \revise{Newer stories were sourced from either accredited data storytelling sites (e.g., Bloomberg) or the Information is Beautiful awards~\cite{infoaward}. We also added data stories that addressed minority domains in the merged corpus.}

\revise{To develop the codebook, which we used to analyze our data story corpus, we consulted with experts in literary media and data storytelling, including published authors.
Through multiple open-ended consultations and discussions, we gained an understanding of their process of character creation and character development.
These insights laid the foundation for our codebook, including identifying a fundamental set of character roles and their effects on the narrative.
We then shared the codebook with two separate groups of data visualization and data storytelling experts, whose feedback helped further translate and contextualize these insights for data storytelling.
In each session, we exchanged our findings and translations from the literary consultations. 
Each session took approximately 2--3 hours.
This process broadened the dimensions of the codebook. For example, we include a new type, real-world photos, for identifiable central characters, to be illustrated in~\autoref{sec:mainandicc}. We also expand the types of \AC{antagonist}, to be elaborated in~\autoref{sec:antagonist}.
The visualization experts and data story practitioners validated the finalized codebook. 
Using the codebook, two researchers independently coded the corpus. They met for three sessions to compare and discuss any mismatches until reaching a consensus.  The finalized codebook and corpus are on our website\revise{~\footnote{https://chaorientdesignds.github.io/}}. 

}

From our analysis, we derive that the role \revise{that} visual elements and visualizations serve in data storytelling falls into two states: (1) \textit{given an existing visualization, how can we adapt it to emphasize and explain a finding} ~\cite{srinivasan2018augmenting} and (2) \textit{given a finding, how can we visualize it}~\cite{shi2020calliope}. 
\revise{Both states share the communication goal of data storytelling; however, they present two different starting points for what we view as characters. The former has an existing character that will be developed and altered to depict the plot, whereas the latter starts from scratch. 
This work delves deeper into the former. We investigate how the 160 data stories would be framed with data characters. We find that the constant involvement of a data character in all the story pieces can support the audience in seeing and tracking continuity in a story.
Future work may investigate concrete strategies for maintaining and presenting the connections between story pieces (i.e., through-line) in a story, including how to elicit an emotional connection to the stories.}


\section{Character Roles \revise{\& Behaviors}}
\label{sec:roles}
In written storytelling~\cite{truby2008anatomy,campbell2008hero}\revise{,} there are many character roles in addition to the main character\revise{, such as a deuteragonist or a love interest}. 
\revise{
In the simplest form of a story, there would be a protagonist, a character that drives the plot forward, and an antagonist, a character that stands in direct opposition to the protagonist~\cite{truby2008anatomy,campbell2008hero}.
}
\revise{For simplicity,} in this work, we only focus on just three essential roles: \MC{main}, \SC{supporting}, and \AC{antagonist} characters.
\revise{
It is documented that a main character need not be the protagonist, as often there are other characters that can advance the story, even the antagonist is included~\cite{campbell2008hero}.
In a data story, the protagonist could be the audience, as they interact with data stories and can be the ones that drive the story plot forward, whereas the main character remains a visual element in the story.
Future work may investigate the nuances and interplay of characters and potentially data character roles.
}

\subsection{Main Character}
\label{sec:mainandicc}
\revise{The core of our data characters is a concept. While concepts are often intangible ideas, let alone visual, this cannot be the case for the \MC{main character (MC)}. After all, data storytelling is a visual medium. The \MC{MC} is the device that the audience can rely on to make sense of and contextualize what is transpiring; therefore, the \MC{MC} should be visually present in the story. More precisely, we need a visual element that can be identified by the audience as the center of the story, to which we refer as an \textbf{identifiable central character (ICC)}.
The ICC becomes the vehicle that visually navigates the audience through the story plot to the conclusion.} 
After reviewing our corpus of data stories, we \revise{find} the visual representation of the \MC{MC} \revise{(i.e., potential ICCs)} ranges from an abstract representation to a real person.
As shown in~\autoref{fig:icc}, we present the common forms we identified from our analysis. 

\revise{
The ICC can be any visual encoding, visualization, or set of visual elements, so long as it is central to communicating the core message.
}
It is up to the author on what visual representation will suit their narrative needs the best.
\revise{For example, in the data story about the U.S. immigration from Cruz et al.~\cite{cruz2018process}, the ICC is a natural metaphor, as shown in~\autoref{fig:icc}b. The story begins with the authors revealing this data-driven natural metaphor without any context; the audience thus may not be able to understand the inherent meaning.}

When the story begins, our \MC{MC} \revise{starts} in the ``ordinary world'' or an expected state\revise{~\cite{campbell2008hero}}.
In this initial state, the audience should be primed with the relevant context and understanding to decode what is visually presented.
The introduction of the character must get across the relevant context, and background information as well as decoding information to relate the visual element to what they mean.
Furthermore, it sets up the motivation and purpose of this character.
It should also introduce doubts, the uncertainty, and begin to challenge the character in its purpose.
This context drives the data character's actions and sets up expectations for the audience.
For example, the U.S. immigration story opens up by introducing a large \MC{tree ring} with a description of how it represents U.S. immigration.
However, the explanation for what the colors mean is initially left unanswered, prompting the audience to scroll onward to learn more about what they are seeing. 

\revise{The behaviors of a character} are driven by its desire to reach a personal goal~\cite{campbell2008hero}. The desire is determined by the author by factoring in their intent, the through-line, and the story pieces. The \revise{lens of the audience} into a story is through the \MC{MC}. The \MC{MC} has a \textbf{desire} that it seeks throughout the course of the story. \revise{For simplicity, }the \revise{other two} characters either aid the \MC{MC} in attaining \revise{its} desire or are in opposition. 
\revise{Those in opposition will bring the \MC{MC} to change and evolve throughout the story, and the audience will learn what those changes mean.}
In the case of~\autoref{fig:icc}b, the \MC{MC} \revise{may} desire to \textit{inform} the audience about population growth and demographic changes over the years in the United States. \revise{As the audience scroll onwards, the authors introduce more characters in the story, illustrated in the following subsection.}

\revise{Although it is ideal for a \MC{MC} to also be the sole identifiable central character, ICC, there are exceptions. 
With other forms of storytelling, technically, there can be identifiable characters that are central to the story (e.g., \textit{The Lord of the Rings}).
It is important to be cautious when having multiple ICCs in a story, as they will compete for the attention of the audience. 
This may result in issues with the delivery of the data storyteller's core message.
Introducing more characters will increase the complexity of the story, but depending on the communicative goal, it can be effective in conveying the intended message.} 

\begin{figure}
    \centering
    \includegraphics[width=\linewidth]{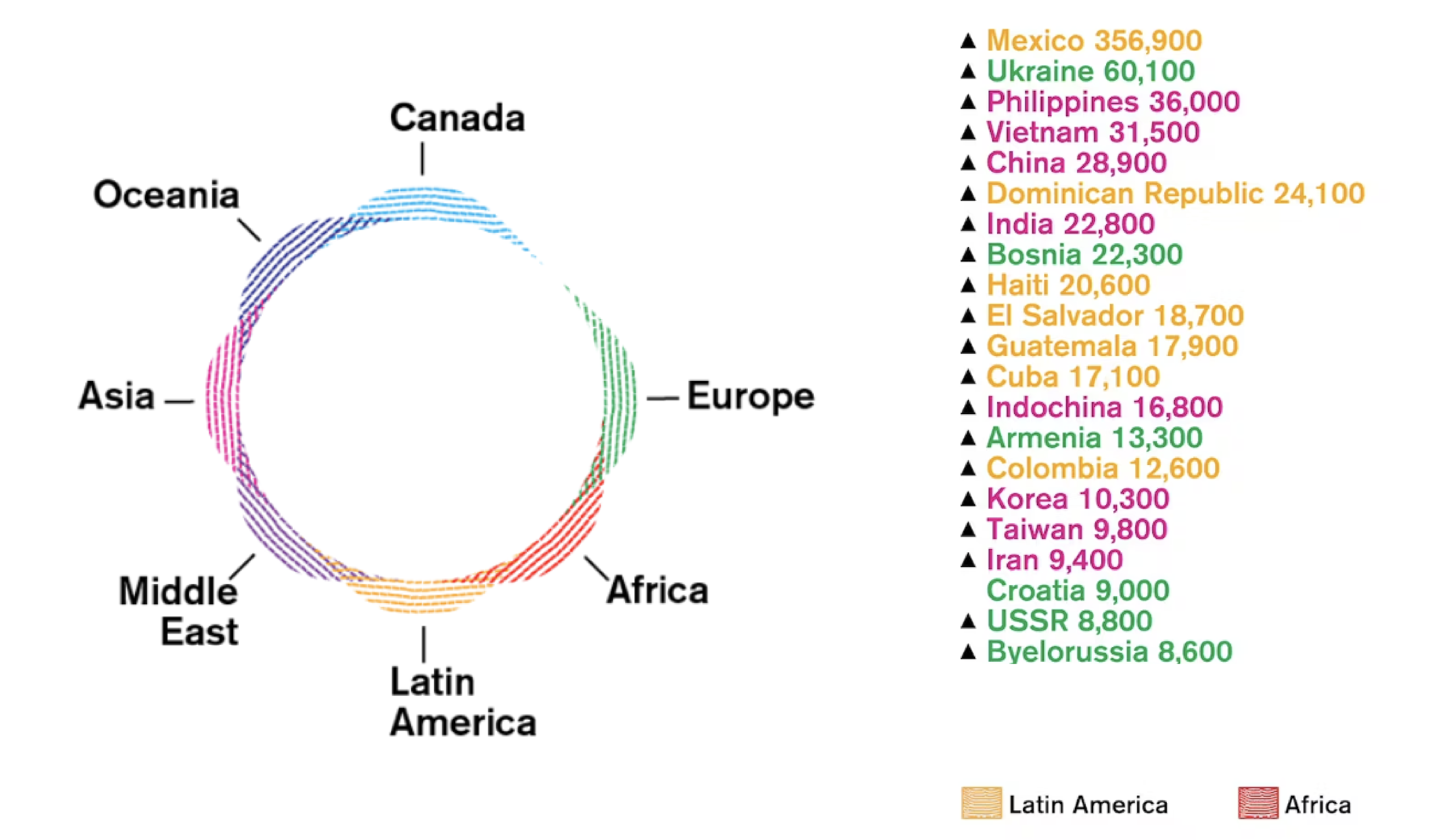}
    \caption{\revise{An example of two \SC{SCs}}. The \MC{MC} is a natural metaphor illustrating the U.S. immigration~\cite{cruz2018process}. The \SC{SCs} of this story \revise{include} a legend and additional details of where immigrants are emigrating from. These \SC{SCs} provide extra information \revise{for the audience} to better understand the \MC{MC}.}
    \label{fig:supporting_character}
\end{figure}

\subsection{Supporting Characters}
A \textbf{\SC{supporting character (SC)}} brings out dimensions of the \MC{MC} and helps push \revise{it} towards \revise{its} desire.
\SC{SCs} must have some relationships with either the \MC{MC} or the \AC{antagonist}, but do not require a relationship among themselves\revise{, as shown in }~\autoref{fig:characterweb}.
For \revise{simplicity and} the sake of \revise{clear story} illustration, \revise{we recommend} not to have a \SC{SC} that supports \revise{both the \MC{MC} and the \AC{antagonist}}; however, it is possible \revise{(e.g., the betrayer in storytelling)}.
The \SC{SC} needs not actively work towards \revise{supporting the \MC{MC} to pursue its desire; rather, }it should never intentionally impede the \MC{MC}.
We found \SC{SCs} in data stories are often \revise{tasked with providing} missing context, extra information, and even alternative representations of the data.

Continuing with \revise{the story of the U.S. immigration}, we left off with the introduction of \revise{the \MC{MC}}, a data-driven natural metaphor of a \MC{tree ring}. As the story advances, \revise{we encounter other} characters that offer \textit{extra information} \revise{and} \textit{alternate representations} to help the audience gain a deeper understanding of the \MC{tree ring}. \revise{As shown in}~\autoref{fig:supporting_character}, we see a \SC{legend} that decodes the \revise{visual} encodings, allowing the audience to understand what \revise{the color} means and infer more insights about the \MC{tree ring}. 
The \MC{tree ring} is also accompanied by a \SC{list of locations}, providing extra information on the geographic origins of individuals immigrating to the U.S.
These devices help the audience understand and interpret our \MC{MC} as the story progresses.

\subsection{Antagonistic Character / Force}
\label{sec:antagonist}
\begin{figure}
    \centering
    \includegraphics[width=\linewidth]{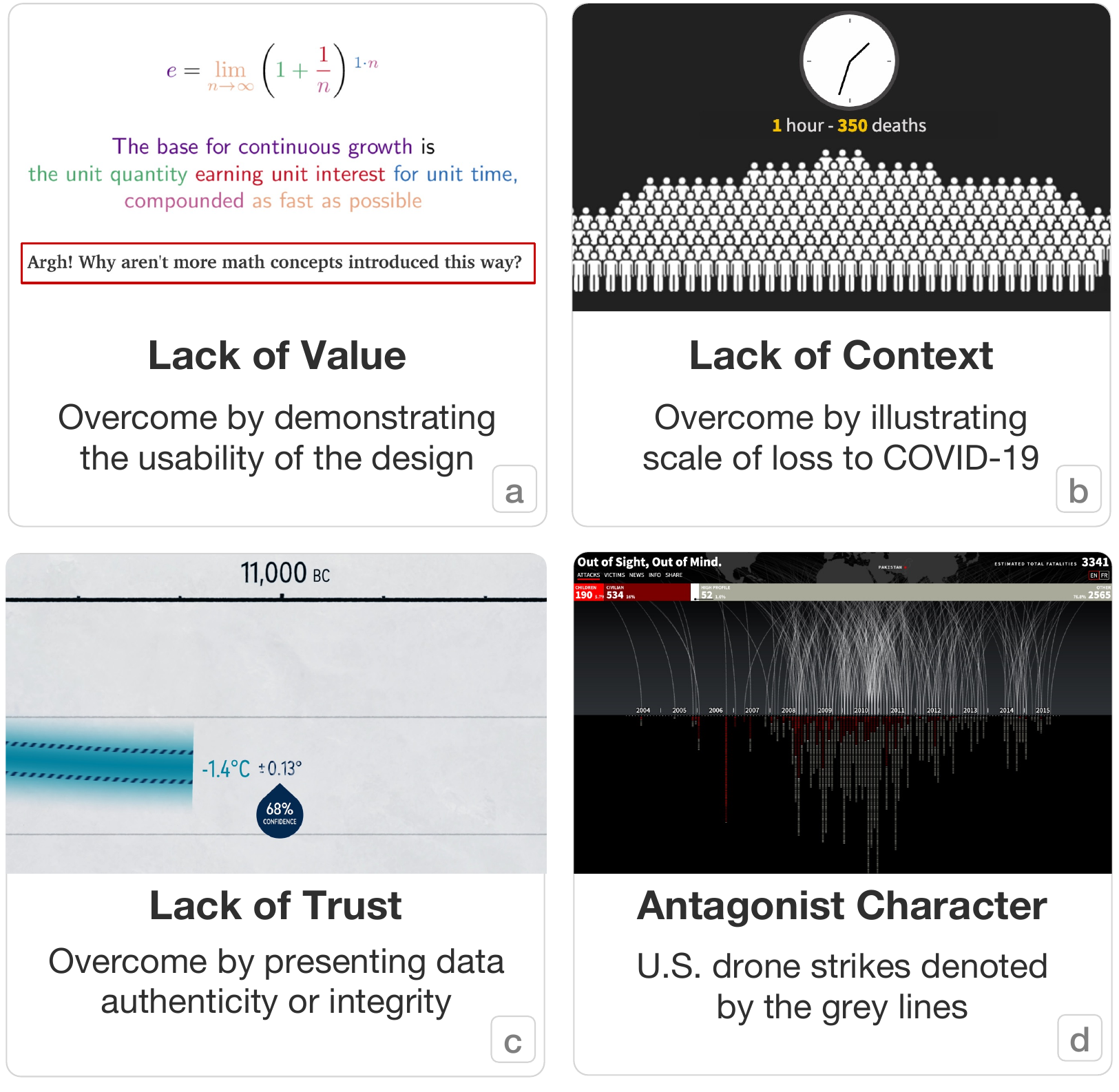}
    \caption{Examples of \AC{Antagonist} Forces (a--c) and Character (d) in DS. (a) \revise{The story about colorized math equations~\cite{azad_2019}} argues \revise{the 
 design} \textbf{value}. (b) The \revise{story ``Pace of Death''~\cite{scarr_2020}} provides the \textbf{context} for how many have died from \revise{COVID}-19. (c) The \revise{video ``Degrees of Uncertainty''~\cite{howsure_2021}} attempts to demonstrate the \textbf{integrity} of the data analysis for climate change. (d) \revise{The story ``Out of Sight, Out of Mind''~\cite{outofsight2004}} depicts the antagonist as drone strikes.}
    \label{fig:antagonist}
\end{figure}

The difference between an \AC{antagonistic force (AF)} and an \AC{antagonist character (AC)} is that an \AC{AF} is an ethereal presence that exists but is not seen directly\revise{, whereas an \AC{AC}} is visually present \revise{in the story}. We found that when the \AC{antagonist} appears as a visual character in data stories, it is often the \MC{MC} driving the story forward. \revise{As shown in~\autoref{fig:antagonist}-(d), there is} a depiction of an \revise{\AC{AC}} in the \revise{data story} ``Out of Sight'' about \revise{the} U.S. drone strikes in Pakistan. The \AC{AC} in this story are the grey lines that symbolize \revise{individual} U.S. drone strikes.

\revise{If there is no \AC{antagonist}, the \MC{MC} will achieve its desire unimpeded. 
In the context of visual data storytelling, the \AC{antagonist} needs to be present to stop the \MC{MC} from pursuing its desire. 
\textit{How can a visualization have a ``villain'' or ``antagonistic force'', and what does that imply?} 
In \revise{our framework}, the \AC{AF} can represent \textit{misunderstandings} and \textit{misconceptions}.}
A misconception is a mistaken belief or having the wrong idea. For example, it is a misconception to hold the belief that the earth is flat.
A misunderstanding is \revise{having different interpretations of a meaning}. As an example, thinking that a rainbow color map implies weather data would be a misunderstanding.
\revise{Throughout the story, these \AC{AFs} or the \AC{AC} prevent the \MC{MC} from achieving its resolution.}
In the context of the U.S. immigration data story, the antagonist is not visually present and therefore is an \AC{antagonistic force}. The force manifests as misconceptions being hurled at the \MC{MC} about a lack of context and understanding of where people immigrated from. 

From our analysis, we find the \AC{antagonist} in data stories often is a force. \revise{We further} identify three forms that these \AC{AFs} assume \revise{to represent} misconceptions and misunderstandings \revise{---} \textit{lack of value}, \textit{lack of context}, and \textit{lack of trust}.

\noindent\textbf{Lack of value.} The \AC{AF} creates external conflicts with the characters in the form of questions, such as ``why is the design meaningful?'' or ``why do microtubules matter?'', driving the story plot to address these questions.
In a data story about colorized math equations\revise{~\cite{azad_2019}, as shown in}~\autoref{fig:antagonist}-(a), the \AC{AF} is constantly questioning the value of the design and its usability. 
As a result, the story attempts to motivate and demonstrate the design is effective.

\noindent\textbf{Lack of context.} The characters face external conflicts due to missing information, \revise{leading to} a misunderstanding or a complete misconception. 
Often this \revise{kind of} conflict is resolved by showing the ``scale'' of a phenomenon (\revise{e.g.,}, the scale of loss, the scale of gun violence, or the scale of climate change).
For instance, as shown in~\autoref{fig:antagonist}-(b), the data story ``Pace of Death'' addresses the \AC{AF} by depicting the number of people who passed away, in a given time interval, due to \revise{COVID}-19. 

\noindent\textbf{Lack of trust.} The \AC{AF} introduces internal conflicts where the characters must prove or refute claims on their integrity (e.g., data authenticity, uncertainty, or credibility).
\revise{In our analysis, we find t}his can be seen in the form of uncertainty visualizations or visualizations that delve into how the ``black box'' of \revise{a machine learning} model behaves.

In all these instances, the \AC{antagonist} is constantly attempting to challenge the \MC{main character} and introducing conflicts to prevent the \MC{MC} from \revise{reaching its desire}.

\subsection{Conflict \& Tension}
\revise{The rationales behind developing and understanding data characters include (1) structuring story content via characters helps filter out irrelevant information and leaves the data storytellers only with the content serving their communicative goal and (2) it unlocks the device of \textit{conflicts}, akin to the debate in the context of science-based works.}
The difference between an explanation and a story is a conflict. 
\revise{Here, we illustrate more on internal and external conflicts.}

\textit{Why do we want conflicts in a science-based story\revise{? W}ouldn't that obfuscate, if not detract from, the \revise{messages} we want to get across?}
\revise{The conflict} is a device that helps the audience understand the motivation of the \revise{story} and can lead to an appreciation for the endeavor or relevancy of what is being communicated.
In other forms of storytelling, \revise{the} conflict is used as a means to bring \revise{the audience} into the story and to form an emotional connection to the characters, such that the audience cares or want the character to succeed in fulfilling its desire.
From our analysis, we find \revise{the} conflict assumes two forms, \revise{\textit{internal} and \textit{external}} conflicts.
Internal conflict \revise{exists} when a character struggles with their own opposing desires or beliefs. It happens within \revise{the character and drives its} development. 
\revise{E}xternal conflict sets a character against something \revise{(e.g., the \AC{antagonist})} or someone \revise{(e.g., the reader)} beyond its control. External forces stand in the way of a character’s motivations and create tension as \revise{it} tries to reach \revise{its} goal.
A majority of the data stories that we review tend to contain external conflicts where the story characters are fighting external forces to either refute claims or provide the missing context to remove a misconception or misunderstanding.
The story resolves once the conflicts are addressed.

\subsection{Data Comic -- ``Something\revise{'}s wrong''}
\revise{To better understand data characters, their roles, and the conflict, we will go through a data comic and address the questions as follows. }
\begin{itemize}[noitemsep,topsep=0pt,parsep=0pt,partopsep=0pt]
    \item \revise{Who is the \MC{MC}?}
    \item \revise{Does the story have ICC? If so, what kind of ICC it is?}
    \item Who are the \SC{SCs}?
    \item What is the \AC{antagonist}? How does it introduce the conflict?
    \item What is the relationship between the plot and the characters?
    \item What is the main theme, \revise{i.e.,} the through-line?
\end{itemize}

\begin{figure}
    \centering
    \includegraphics[width=\linewidth]{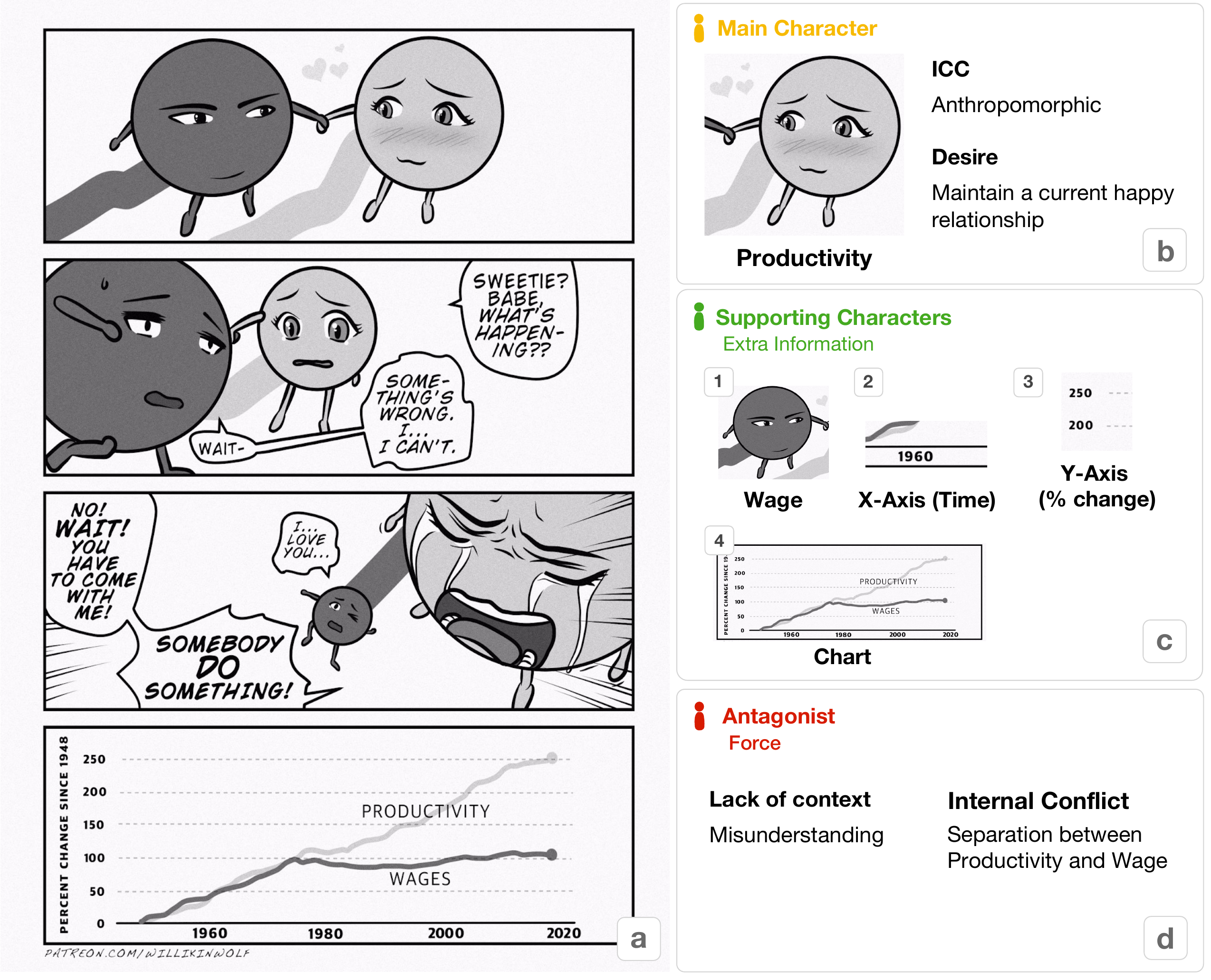}
    \caption{\revise{The data comic ``Something's wrong'', created by Willikin Woolf~\cite{WillikinWoolf} and included in the} data comic gallery~\cite{datacomicgallery}. (a) Original data comic (b) The \MC{MC} -- \MC{productivity} (c) The \SC{SCs} -- (c1) \SC{wage}, (c2) \SC{x-axis} with the year, (c3) \SC{y-axis} with percentage changes, and (c4) the line \SC{chart} with the same colors as \MC{productivity} and \SC{wage}.  (d) The antagonist is an unaddressed force (\AC{AF}), i.e., the reason behind the separation.}
    \label{fig:data_comic}
\end{figure}
\revise{Data comic~\cite{bach2018design}, inspired by the visual language of comics, is a rising and popular genre for presenting information effectively.}
We will \revise{analyze} a data comic \textit{Something\revise{'}s wrong} by Willikin Woolf~\cite{WillikinWoolf}\revise{, featured on the data comic gallery curated by Bach et al.~\cite{datacomicgallery}.}
This story depicts the relationship between two \revise{identifiable comic} characters.
The \MC{MC} \revise{is likely} the ``productivity'' character, \revise{as shown in~\autoref{fig:data_comic}-(b)}.
\revise{This is because from~\autoref{fig:data_comic}-(a) panel 3, we see the panel focuses more on the despondent \MC{productivity}, rather than the ``wage'' character.} 
We may infer that the desire of \MC{productivity} is to maintain \revise{the} current happy relationship with ``wage'', as seen in~\autoref{fig:data_comic}-(a) panel 1.
\revise{The visual representation of the \MC{MC} (i.e., ICC) is anthropomorphism.}

\revise{The \SC{SCs} are wage, the x- and y-axes, and the chart, indicated in~\autoref{fig:data_comic}-(c1) to (c4).
As shown in~\autoref{fig:data_comic}-(a) panels 1-3, \SC{wage} seems eager to stay with \MC{productivity}.
Meanwhile, the \SC{axes} and the \SC{chart} are only introduced in the last panel.}
The \SC{axes} help us understand more about \SC{wage} and \MC{productivity} \revise{for providing} the additional context of their percentage change \revise{over time, from similar trends to drastic separation.}
\revise{This is because the colors of both characters correspond to the line colors in the \SC{chart}.}

The \AC{antagonist}, in this story, would be the reason behind this separation.
\revise{Visually absent in the story, this \AC{AF} remains unaddressed or unexplained either, at least on this page of the comic.}
The \AC{AF} introduces the conflict to \MC{productivity} \revise{in the form of a separation} from \SC{wage}. 
The through-line between all these events is to explain the relationship between \SC{wage} and \MC{productivity}.
A possible theme could be to persuade the audience into thinking that these two characters should not drift apart and have a linear relationship.

\begin{figure}
    \centering
    \includegraphics[width=0.82\linewidth]{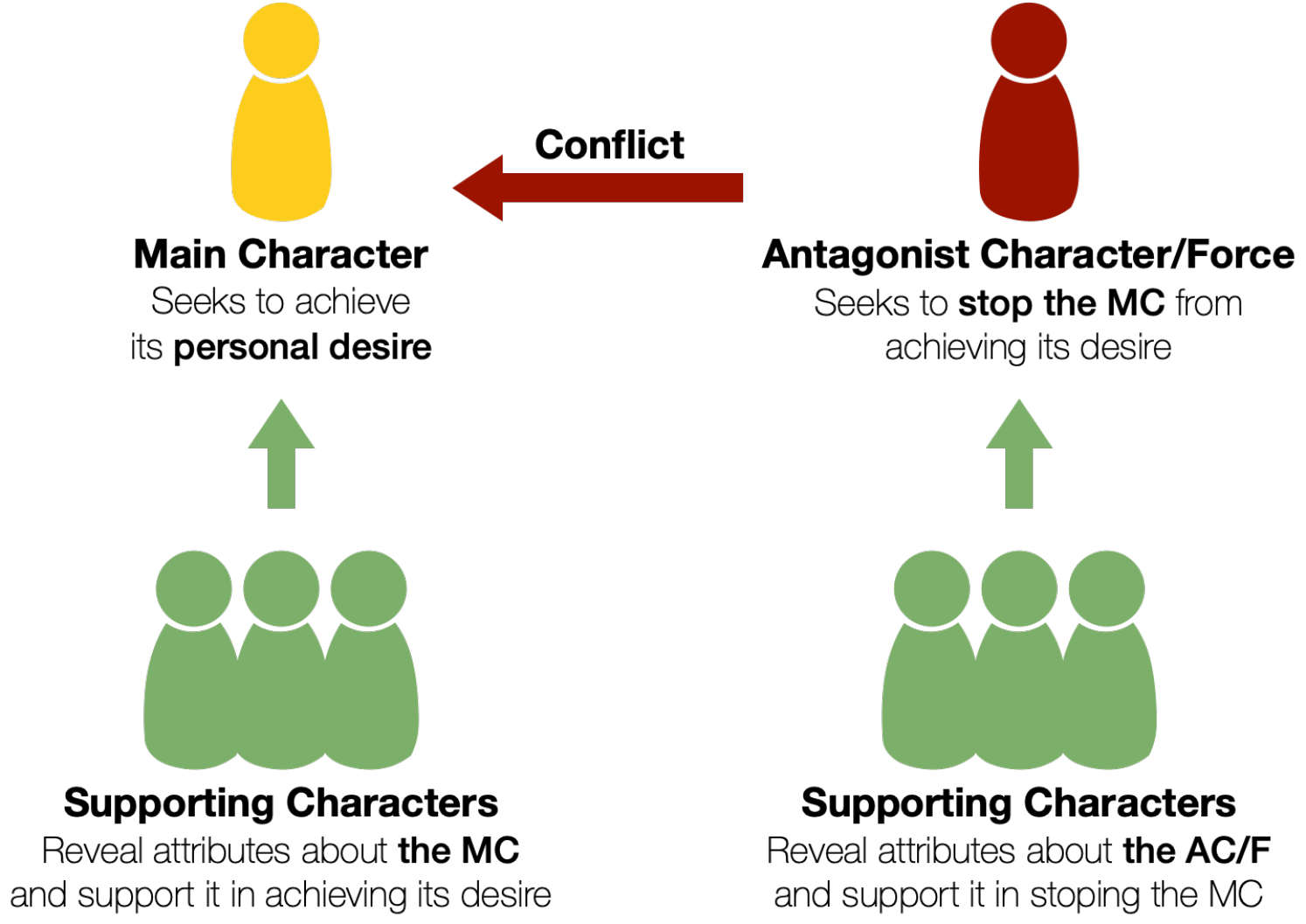}
    \caption{Character Web. The relationships between \MC{main}, \SC{supporting}, and \AC{antagonist} characters. The main character (MC) has a desire and is the focus of the story. The antagonist force (AF) tries to prevent the MC from achieving its desire. Supporting characters (SC) reveal dimensions about either the MC or AC/F. }
    \label{fig:characterweb}
\end{figure}

\begin{figure*}[h]
    \centering
    \includegraphics[width=\linewidth]{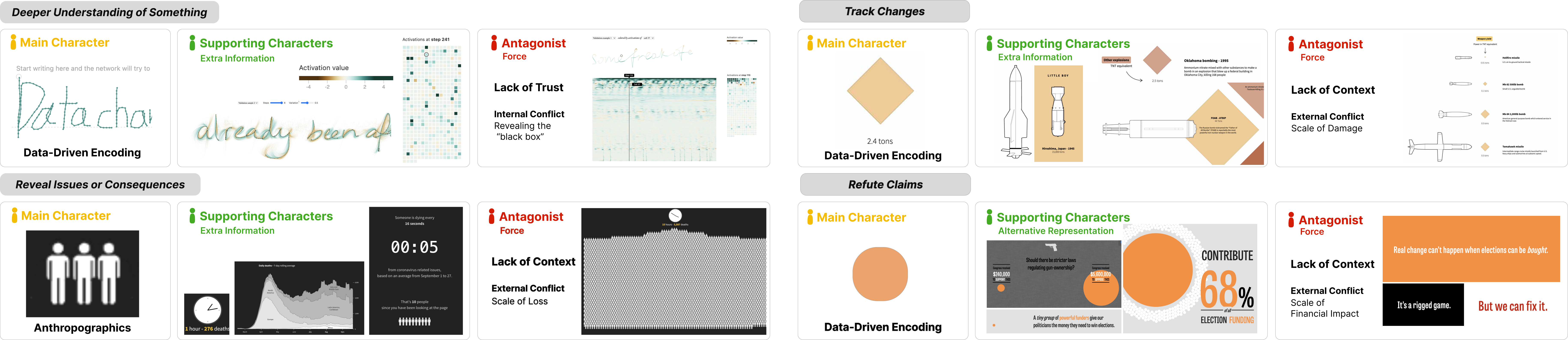}
    \caption{ Plot Types and Data Characters. For four of the five plot types, we illustrate through existing data stories how data characters appear and behave. The first column represents the \MC{main character} followed by \SC{supporting} and \AC{antagonistic forces}. Image sources:~\cite{handwriting_2016, blast_2020, scarr_2020, money_2014}.}
    \label{fig:plot_char}
\end{figure*}

\section{Character Oriented Design Space}
\label{sec:designspace}
\revise{From our framework and analysis of data stories, we have unpacked a data character, identified common representations of data characters, defined properties of basic data character roles, and contextualized conflict in data storytelling.
In this section, we describe how data characters, as defined in this work, can be developed as well as how they could fit into the broader data storytelling space.
When we refer to this space as \textit{character-oriented design}, we intend for it to be used as a guide for helping data storytellers in developing or identifying characters in their stories. }

\subsection{Character Motivation}
\revise{To begin, we first outline our process for character creation, which was also informed by the aforementioned discussions with experts.}
We consider a \textbf{character} as the lens for our audience to view the abstract, which often is a complex concept, we intend to convey. How a character is defined determines both the presentation of the data \revise{and the story focus.}
A data character is a derivation of a character, thus it must adhere to some core storytelling principles: (1) a data character must serve a narrative goal and (2) a data character must have a \textbf{desire}. 
We define \textbf{desire} as the rationale that motivates a character \revise{pursuing} a goal. 
\revise{By assigning a desire to a data character, we should ensure the behaviors of the character are associated and align with this base desire, throughout the data story. 
That is, the character is consistent in its behaviors as it pertains to achieving this desire.}  
Contextualized for data storytelling, this desire is the communication of some scientific concept, linked to an existing dataset with varying \revise{intentions}. 
\revise{We offer the following steps as a suggestion for developing a data character and their \textbf{desire}.} 
\begin{enumerate}[noitemsep,topsep=2pt,parsep=0pt,partopsep=0pt]
    \item Identify the meta-concept (through-line).
    \item Distill the concept into smaller more distinguishable aspects.
    \item Relate these aspects to the data (i.e., via examples or non-examples).
    \item Rank which aspects best illustrate the concept.
    \item Select the visual representation.
\end{enumerate} 

To not lose sight of the original intention of the story\revise{, and to track whether our characters are in line with their desires,} it is good to make use of a through-line (~\autoref{table:storytelling_defs}). 
\revise{As discussed in~\autoref{sec:datastoryprocess}, data stories can be composed of many plots and subplots, steering the original narrative in many directions.
A through-line in data storytelling would be the core concept that ties these subplots together, providing guidance and relating back to the original narrative.}
Identifying a through-line requires the assessment of the story pieces for what common theme or over-encompassing plot could link these \revise{pieces} all together. If we have five disjoint story pieces, then the through line has to be the commonality among the pieces. Alternatively, if we cannot find a common link between these pieces, we \revise{should then consider re-evaluating} what we seek to convey. 
In our case, for data storytelling, this commonality is a scientific concept that each individual piece relates to. 
We can utilize the seven \revise{story} plot types~\cite{ojo2018patterns} \revise{along} with the seven genres~\cite{segel2010narrative} to identify a central theme, through-line, for our story. 
As for the situations where certain story pieces address disparate concepts, one \revise{of them} must be prioritized. The \revise{remaining} groups should either be omitted~\cite{hullman2011visualization} or be part of another story. 

\vspace{2pt}
\noindent \textbf{Premise \& Through-Line:} When organizing a data story, the first step is to \revise{identify} what is the through-line between our story pieces. \revise{Namely, \textit{what is the meta-concept that ties the messages we want to convey?}}  If we know our through-line, we \revise{may} understand how our characters relate to the plot. 
Then, we can create a premise, or a small road map, of the entire story. A premise is the simplest expression of the story being told and presents a sense of the \MC{MC} as well as the outcome of the story. 

\vspace{2pt}
\noindent \textbf{Possibilities:} Once we have an understanding of the premise and through-line, we should identify what is possible in this premise. What types of genres do we have access to, what structures can we make use of, and what setting will it take place in? 

\vspace{2pt}
\noindent \textbf{Challenges + Problems:} \revise{Identifying the unique challenges and problems ties} to conveying the content of our data story. These problems could be the audience \revise{lacking} familiarity with what is being expressed, the subject matter is too abstract, there are difficulties simplifying the science, and so on. We need to note what the main misunderstanding or misconceptions about the story content are. 

After answering these points, we have developed a foundation for our \MC{main character} and \SC{supporting characters}. The next challenge is determining what should be our \MC{main character} and how to develop this from our through-line and \AC{antagonist}. 

\subsection{Character Creation}
Naively, a data character inherits the properties of a data-driven visual element
It would likely contain a set of attributes and behaviors that relate to explaining the concept. A character can be a lens into \revise{a meta-concept or scientific domain}. 
Initially, a character represents an idea in a world full of misconceptions, misunderstandings, and uncertainty. The role of a character in the story is often to explore its relationship to this perceived \revise{world} and attempt to overcome misconceptions\revise{, asserting} its place. Contextualized for communicating scientific information, a data character would be \revise{born} from data or namely analytics. As \revise{discussed in~\autoref{sec:mainandicc}}, the motivation of a character is its desire
\revise{, which would} be the core \revise{message} to get across. 
\revise{Thus, it is important to understand what concepts represent a data character.}
Depending on the complexity of the core concept\revise{, we often may rely on} multiple characters to best convey it. This is due to concepts having many dimensions that should be viewed and expressed with varying lenses. The concept we prioritize and give the focal lens would be the basis of a \MC{main character}.
Collections of smaller aspects that exemplify and illustrate concepts as a collection would result in \SC{supporting characters}. 
Aspects from the data that\revise{ may challenge the concept}  (i.e., contradictions) can lead to potential \AC{antagonists}. 

\revise{We suggest the best starting point is the meta concept} that contains the smaller \revise{sub-concepts, i.e.,} the through-line.
\revise{Once we identify the meta concept, we may} create sub-groups of the findings to express more distinguishable aspects of the \revise{meta} concept.
\revise{We then can investigate how each sub-group illustrates the aspects of the meta concept. Precisely, we inspect whether each sub-group serves a direct example, or perhaps a non-example (i.e., contradiction), of the meta concept.}
\revise{These relations allow us to rank each sub-group based on their capability of demonstrating these aspects.}
Some \revise{sub-groups} are stand-alone, some aspects may show details of phenomena in others, while some may introduce doubt or uncertainty. This gives us a \revise{basis} for the \MC{main}, \SC{supporting}, and \AC{antagonistic} characters. \revise{This can also be the point where data storytellers begin to design the visual elements that will portray these characters.}
\section{Characters \revise{in Story Plots}}
\label{sec:relationships}
\revise{
From our analysis of the corpus, we identify several patterns in how characters often behave in the stories.
In this section, we present four concrete story examples to describe how the relationships of the character roles, as shown in~\autoref{fig:characterweb}, could be contextualized in the story plots.
Here, we first introduce the initial seven story plots for data stories, identified by Ojo and Heravi~\cite{ojo2018patterns}:
}
\begin{enumerate}[noitemsep,topsep=2pt,parsep=0pt,partopsep=0pt]
    \item Refute claims.
    \item Reveal unintended consequences.
    \item Reveal information of personal interest.
    \item Enable deeper understanding of a phenomenon.
    \item Reveal anomalies and deficiencies in systems.
    \item Reveal information about an entity in increasing levels of detail.
    \item Track changes in systems.
\end{enumerate}

\vspace*{0.05in}
We \revise{may} consider ``phenomenon'', ``entity'', and ``system'' as exchangeable main characters, depending on the communicative goal.
\revise{
Consequently, we collapse the second and the fifth plot types into \textit{reveal anomalies, issues, or unintended consequences}.
We merge the fourth and the sixth plot types into \textit{deeper understanding of something}, as these two story plots explore a character (e.g. phenomenon or entity) in depth.
By factoring in the character roles, we integrate these seven story plots into the following five types:
}
\begin{enumerate}[noitemsep,topsep=2pt,parsep=0pt,partopsep=0pt]
    \item Refute claims.
    \item Reveal anomalies, issues, or unintended consequences.
    \item Reveal information of personal interest.
    \item Deeper understanding of something.
    \item Track changes.
\end{enumerate}

\vspace{2pt}
\revise{
In the remaining section, we describe how data characters may be woven into each of these story plots, including a generic through-line, the desire of \MC{MC}, \AC{AF}, and a possible premise of the story plot.
We focus on \AC{AF} as we find it to be the majority of \AC{AF/C} in our corpus, whereas \AC{AC} is rather intuitive due to its visibility.
While these examples do not represent the only way to construct such a story plot, they illustrate how we may pair characters and story plots together to organize the story.
}

\vspace{2pt}
\noindent\textbf{Refute claims.} 
 %
The through-line can be to persuade the audience into believing the inverse of the claim.
The \MC{MC} desires only to prove \revise{the inverse of the claim}. 
The \AC{AF} may take the form of misconceptions or misunderstandings that support the claim. \revise{Consequently, the conflict, when resolved by the \MC{MC}, helps refute the claim.}
\textit{Premise:} the \MC{MC} heads towards the state\revise{,} where the world is opposite of the claim. The \AC{AF} causes conflicts to prevent the \MC{MC} from reaching its desire. By overcoming the conflict, the \MC{MC} fulfills its desire and refutes the claim.

For example, we examine the story, \textit{Money Wins Elections}, as shown in \autoref{fig:plot_char}.
The through-line is \revise{a claim that ``money wins elections'', with other concepts addressing corruption in the U.S. government.}
\revise{
The \MC{MC} is a \MC{point mark} that represents a vote, where its size encodes financial investments. 
We may infer the claim the \MC{MC} seeks to refute is that ``all votes are equal'' or that ``nothing can decide election outcomes'', as the \MC{MC} desires to persuade the audience that money can buy election results. 
The \AC{AF} is the misconception that causes external conflicts. The \MC{MC} resolve the conflict by demonstrating that the scale of financial investment can influence the election outcome.
}

\vspace{2pt}
\noindent\textbf{Reveal anomalies, issues, and unintended consequences.}
The through-line \revise{is theme-dependent and affected by} the disposition of the author towards the consequences or unexpected events.
The desire of the \MC{MC} is to maintain the world state as expected.
The \AC{AF} introduces an action or event that creates conflict \revise{with the world \MC{MC} expected}.
\revise{While the \MC{MC} addresses this conflict, the result causes the \MC{MC} to deviate from the direction it expected.}

Referring to~\autoref{fig:plot_char}, we see an instance of this type of story,  \textit{The Pace of Death}.
The through-line is \revise{COVID-19}, with a sub-concept of mortality rate.
The \MC{MC} is a person depicted by an anthropographic \MC{icon}. 
There are several \SC{SCs} providing extra information about the person, such as a \SC{clock} illustrating how many people pass away after a period of time. 
The \MC{MC} is initially under the impression that the mortality rate is not high. As the story advances, the \AC{AF} reveals what the true mortality rate looks like.
\revise{One of the \SC{SCs} is a \SC{timer} that reveals the scale of loss that has been incurred while the audience reads this story; thus, we may infer the primary intent of the author is to terrorize the audience to call attention to the severity of COVID-19.}

\vspace{2pt}
\noindent\textbf{Reveal information of personal interest.} 
The through-line is often to inform the audience about some information.
The desire of \MC{MC} is to showcase the personal interest, \revise{which is the author's in this case~\cite{ojo2018patterns}}, whereas the \AC{AF} is to dismiss the value.
In this story plot, the \MC{MC} tries to bring attention to a topic and motivate its value. The \AC{AF} introduces conflicts by casting doubt on the topic's value. By overcoming the conflict, the \MC{MC} persists in showing why the topic is interesting. 

\vspace{2pt}
\noindent\textbf{Deeper understanding of something.} 
The through-line is to explain something (e.g., a phenomenon) to the audience.
The desire of \MC{MC} is to explain the mechanisms behind the topic as simply as possible, whereas the \AC{AF} are misunderstandings about this topic.
The \MC{MC} attempts to explain to the audience, however, the \AC{AF}  introduces hurdles based on misunderstandings. To overcome the conflict, the \MC{MC} must unpack the content further, until the author feels the story goal is achieved.

In the story, \textit{Four Experiments in Handwriting with a Neural Network}, the \MC{MC} is the \MC{handwritten phrase} of the user, as shown in~\autoref{fig:plot_char}. The \MC{MC} seeks to better understand \revise{how a neural network learns to the handwriting style}. The through-line is about neural networks with sub-concepts in handwriting and generative models. \revise{The \AC{AF} comes in the disguise of an internal conflict, the black box nature of neural networks}. 
To overcome this conflict, the \MC{MC} unpacks the black box\revise{, the layers of the neural network}. 

\vspace{2pt}
\noindent\textbf{Track changes.} 
The through-line is often to inform the audience of changes in a character (e.g., an entity or a system).
The is the same for the desire of \MC{MC}.
\revise{However, \AC{AF} presents obstacles to MC in the attempt of causing changes.}
In this story plot, the \MC{MC} is trying to stay ``unchanged'' but is presented with obstacles, resulting in conflicts. To overcome these conflicts, the MC must change.

This story plot can be seen in the data story, \textit{How powerful was the Beirut blast?}, as shown in~\autoref{fig:plot_char}. 
The through-line is the devastation that an explosion causes. 
The \MC{MC} is a data-driven \MC{diamond mark} that represents an explosion measured in TNT \revise{equivalent} (e.g., it starts at 0.01 tons). As the story progresses, the \MC{mark} grows larger and larger, \revise{and the narrative changes} to convey how different blasts compare. The \AC{AF} for this story appears as a lack of context, a misconception of the magnitude of how devastating various bombs or explosive accidents are. The \MC{MC} overcomes this conflict by comparing these incidents.



\section{Discussion}
\revise{The goal of our work} is to illustrate how characters can be utilized to frame abstractions and communicate insights or findings through the story plot. \revise{Through our framework}, we have identified specific features, relationships, and roles of data characters. From these findings, we describe a space for developing data characters and applying them in data stories, which we refer to as a character-oriented design space. 
Within this space, \revise{we hope to motivate data storytellers to view visual elements} as characters with narrative goals, rather than data-bound abstractions to be explained. By treating visual elements as characters, \revise{we organize} visual element(s) into those that are driving the story (\MC{MC}), supporting the story (\SC{SC}), and those \revise{contribute via contradiction (\AC{AC/F})}.
\revise{We further investigate how data characters weave into representative story plots for data stories.}
\revise{The scope of this work is to offer a framework for understanding data characters and a design space to serve as the foundation for developing data characters.}
Our work introduces new considerations for the data storyteller: (1) they must develop the characters to best tell their story, (2) the roles (i.e., \MC{MC}, \SC{SC}, \AC{AC/F}) for their characters, and (3) the number of characters needed to tell their story.
\revise{In this section, we discuss the audience as data characters and character roles that are specific to data stories.}

\noindent\textbf{Audience as a character.}
There are many character \revise{roles outside the ones discussed in this work}, and stories can \revise{utilize these roles in a myriad of ways} that we did not address. 
\revise{
When data stories are interactive, it becomes more ``reader-driven''  as the audience is now a part of the story, often the protagonist~\cite{segel2010narrative}.
A protagonist is defined as the character that moves the story plot forward. 
In this role, the audience is not always visually represented as a character, they may control the \MC{main character} and even at times could be an \AC{antagonistic force}, but may not be an ICC themselves.
For example, the audience may play the role of the ``devil's advocate'' towards either the \MC{main character} or the \AC{antagonist}.
In some instances, the audience can be visible and represented as an avatar or other ICC depiction, as shown in~\autoref{fig:icc}.
The audience can also serve as secondary key characters, known as a deuteragonist~\cite{truby2008anatomy,campbell2008hero}, they can play a role akin to either the \MC{main character}, \AC{antagonist}, or even a neutral agent.}


\noindent\textbf{Data story specific roles.}
\revise{A potential direction for future work would be identifying character roles unique to data storytelling. A method that could be applied to identify such roles} is character archetypes, as shown in~\autoref{tab:character_defs}, which are templates for generic characters based on certain types of behaviors and patterns. 
This device can give the storyteller guidance on the specific nature and behaviors that the \MC{MC}, \SC{SC}, or \AC{AC/F} will take on. Archetypes differ in other storytelling \revise{media} as they focus on types of people and the human condition. In novels and screenplays~\cite{truby2008anatomy,campbell2008hero,field2005screenplay}, some examples of archetypes \revise{include}: father, queen, mentor, warrior, and lover. Often these archetypes come with strengths, inherent weaknesses, and understandable relationships to help build a story. 
\revise{However, often it is not} a focus of data stories to discuss or explore the human condition. 

From our analysis, we suggest a starting place for character archetypes \revise{would be} systems, anomalies, entities, and phenomena. 
The properties of the data can indicate some behaviors. For example, hierarchy suggests depth, temporality may imply change, and spatiality could imply closeness or bonds. 
\revise{We may extend this thinking by taking data types into account, such as nominal, categorical, numerical, and their pairings.}
Multiple numerical datasets could contrast with one another. \revise{We discuss three potential data character archetypes; \textit{overview}, \textit{parental}, and \textit{cluster}}.

The archetype of the \textit{overview} would be a character that knows the broad picture of the relationships between other characters, but need not give an opinion, similar to an observer. 
An example can be a visual analytic system for managing production lines.
With stories centered around prediction, we can pull out a \textit{parental} archetype, where the parent has multiple attributes presented as a formula to describe their child.
The story could be interested in the parent-child relationship, what is the relationship's strength, how the child affects the parent, how external negative factors that the child faces also impact the parent, etc.
This sort of archetype could be useful for explainable AI.
Another potential character archetype would be \textit{clusters}. A pairing of numerical and categorical can result in clusters.
They are commonly used in visualization to denote relationships and close associations.
\revise{The properties of a cluster can reveal commonalities, uniqueness}, and how they are affected by the change (e.g., does the group stay or split).

While there are many character roles identified in other forms of storytelling, it is unclear how relevant these are to data stories.
\revise{This work lays the foundation by providing a fundamental set of data character roles (i.e., \MC{MC}, \SC{SC}, \AC{AC/F}), investigates how they weave into data stories, and discusses how the space of data characters could be expanded.}
\revise{We hope this work presents an opportunity within the space of data characters} to extend past the roles identified in this work. 

\begin{table}[t]
\small
\centering
\begin{tabular}{p{\dimexpr 0.37\linewidth-2\tabcolsep} p{\dimexpr 0.63\linewidth-2\tabcolsep}} \hline
    \toprule
    \textbf{Character Terminology} & \textbf{Description} \\
    
    \arrayrulecolor{black!30}\midrule
    
    \textbf{Character} & A visual entity that influences itself and others and serves a narrative purpose. \\

    Desire & 
    The rationale that motivates a character's actions. \\

    Conflict &
    Arises when a character while pursuing their desire faces an obstacle. The pressure that is applied to the \MC{MC} forces change. \\
    
    \arrayrulecolor{black!30}\midrule
    
    Archetypes &
    Patterns within an entity; the behaviors they exhibit that are essential in how they interact with others. \\
    
    \MC{Main Character (MC)} &
    \MC{MC} has the central desire and serves to contextualize or introduce the story/domain. \\

    \SC{Supporting Character (SC)} &
    This character role complements either the \MC{MC} or \AC{AC}. They provide a means to see more depth about the \MC{MC} or \AC{AC}. The desire of \SC{SC} can align with either the \AC{AC} or \MC{MC}, but will not go against it. \\
    
    \AC{Antagonist Character or Force (AC/F)} &
    \AC{AC} seeks to prevent the \MC{MC} from reaching their desire. It causes conflict with \MC{MC}. \revise{The story plays out when the conflict is resolved}. \\
    
    \bottomrule
\end{tabular}
\caption{Character-specific storytelling terminology. These terms and their mappings were derived from a breadth of visual and written storytelling literature~\cite{truby2008anatomy,field2005screenplay,fink2014dramatic,forster2010aspects,campbell2008hero}.
} 
\label{tab:character_defs}
\end{table}

\section{Conclusion \& Future Work}
Our goal with storytelling is to engage the audience while preserving the scientific integrity of the content. A data story should provide an entry point for the audience, a cohesive plot, and a cast of characters to lead the audience along to where the storyteller intends. As storytellers, we want the audience to create an emotional connection with the story and leave with the intended message. By exploring the role data characters play, we believe characters present the pathway to this goal. 

We review 160 data stories and identify features of data characters, the roles they assume, types of antagonists in data stories, types of conflicts for data stories, and the relationships data characters have among one 
another, and \revise{offer a framework for data characters and design space for developing characters and applying them in a data story.} From the perspective of a data storyteller, we show the role of characters and their importance in data storytelling with consideration of where character design should occur. 
We introduce the idea of an identifiable central character (ICC) as a device that data storytellers can use to select their \MC{MC}, illustrated through our case studies, the relationships between the \MC{MC}, \SC{SC}, and \AC{AC/F} in the context of data stories, and provide an outline for how characters can be woven with five types of data story plots. 

For future work, we suggest exploring general patterns in common data stories in terms of \revise{broadening character roles}. We believe, by providing a discussion on the relationships between characters and data-driven visual entities, we can achieve a more consistent language among designers. Stories require a through-line to connect all the story pieces together cohesively. However, these connections should remain clear to the audience. This would require some identifiable central character or characters for the audience to contextualize the presented information and latch on to as the story unfolds. The relationship between the plot and the story is not complete without the characters. The effective weaving of the two gives us the story. Our desire with this manuscript is for the readers to view and think in terms of characters when creating either a data story, narrative, or explanatory visualizations.

\section*{Figure Credits}
\small
\Cref{fig:ds_process} is a partial recreation of Fig.\ 2 from \cite{lee2015more}, which is in the public domain.

\noindent\Cref{fig:icc} (a) Image credit: Neil Halloran, 2021, How Sure Are Climate Scientists, Really? As seen in \small\url{https://www.youtube.com/watch?v=R7FAAfK78_M}. (b) as seen in Fig.\ 9 from \cite{cruz2018process}, creative commons license, an updated graphic can be seen in \small\url{https://web.northeastern.edu/naturalizing-immigration-dataviz/}. (c) worldcupballs.info and Adidas, 2014, The World’s Ball, as seen in \small\url{https://www.nytimes.com/interactive/2014/06/13/sports/worldcup/world-cup-balls.html}. (d) Manas Sharma, Simon Scarr, and Gurman Bhatia, 2020, The pace of death,  as seen in \small\url{https://www.reuters.com/graphics/HEALTH-CORONAVIRUS/DEATHS/xlbpgobgapq/index.html}. (e)  Willikin Wolf, 2019, Something's Wrong, as seen in \small\url{https://twitter.com/WillikinWolf/status/1176006515968241665}. (f) as seen in Fig.\ 10 from \cite{zhao2015data}. (g) Gapminder.org, creative commons license, as seen in \small\url{https://www.gapminder.org/answers/will-saving-poor-children-lead-to-overpopulation/}.

\noindent\Cref{fig:supporting_character} Image credit: Pedro Cruz, John Wihbey, Avni Ghael, and Felipe Shibuya, as seen in \small\url{https://web.northeastern.edu/naturalizing-immigration-dataviz/}.

\noindent\Cref{fig:antagonist} (a) Kalid Azad, 2019, Colorized math equations, as seen in \small\url{https://betterexplained.com/articles/colorized-math-equations/}. (b) Manas Sharma, Simon Scarr, and Gurman Bhatia, 2020, The pace of death,  as seen in \small\url{https://www.reuters.com/graphics/HEALTH-CORONAVIRUS/DEATHS/xlbpgobgapq/index.html}. (c) Neil Halloran, 2021, How Sure Are Climate Scientists, Really? As seen in \small\url{https://www.youtube.com/watch?v=R7FAAfK78_M}. (d) Pitch Interactive, 2004, Out of sight, out of mind, as seen in \small\url{https://drones.pitchinteractive.com/}.

\noindent\Cref{fig:data_comic} Image credit: Willikin Wolf, 2019, Something's Wrong, as seen in \small\url{https://twitter.com/WillikinWolf/status/1176006515968241665}.

\noindent\Cref{fig:plot_char} (a) Shan Carter, David Ha, Ian Johnson, Chris Olah, 2016, as seen in \small\url{https://distill.pub/2016/handwriting/}. (b) Marco Hernandez, Simon Scarr, 2020,  How powerful was the Beirut blast?, as seen in \small\url{https://graphics.reuters.com/LEBANON-SECURITY/BLAST/yzdpxnmqbpx/index.html}. (c) Manas Sharma, Simon Scarr, and Gurman Bhatia, 2020, The pace of death,  as seen in \small\url{https://www.reuters.com/graphics/HEALTH-CORONAVIRUS/DEATHS/xlbpgobgapq/index.html}. (d) Tony Chu, 2014, Money Wins Elections, as seen in \small\url{http://letsfreecongress.org/}.

\acknowledgments{
     The authors wish to thank NorCal Vis for their insights and discussions.
     A special thanks to Spencer Russell Smith for his consultations and input on storytelling.
     We wish to thank all the reviewers and their thoughtful feedback and suggestions throughout the process of developing this manuscript.
}

\bibliographystyle{abbrv-doi-hyperref}

\bibliography{template}

\begin{thebibliography}{10}

\bibitem{amini2015understanding}
F.~Amini, N.~Henry~Riche, B.~Lee, C.~Hurter, and P.~Irani.
\newblock Understanding data videos: Looking at narrative visualization through
  the cinematography lens.
\newblock In {\em Proceedings of ACM Conference on Human Factors in Computing
  Systems}, pp. 1459--1468. ACM, New York, NY, 2015.
  \href{https://doi.org/10.1145/2702123.2702431}
{doi: {{%
10\hspace{.1pt}\discretionary{.}{%
}{.}\hspace{.4pt}1145\discretionary{/}{%
}{/}2702123\hspace{.1pt}\discretionary{.}{%
}{.}\hspace{.4pt}2702431}}}


\bibitem{amini2016authoring}
F.~Amini, N.~H. Riche, B.~Lee, A.~Monroy-Hernandez, and P.~Irani.
\newblock Authoring data-driven videos with dataclips.
\newblock {\em IEEE Transactions on Visualization and Computer Graphics},
  23(1):501--510, 2016. \href{https://doi.org/10.1109/TVCG.2016.2598647}
{doi: {{%
10\hspace{.1pt}\discretionary{.}{%
}{.}\hspace{.4pt}1109\discretionary{/}{%
}{/}TVCG\hspace{.1pt}\discretionary{.}{%
}{.}\hspace{.4pt}2016\hspace{.1pt}\discretionary{.}{%
}{.}\hspace{.4pt}2598647}}}


\bibitem{azad_2019}
K.~Azad.
\newblock Colorized math equations.
\newblock
  \small\url{https://betterexplained.com/articles/colorized-math-equations/},
  2019.

\bibitem{datacomicgallery}
B.~Bach.
\newblock \small\url{https://datacomics.github.io/}.
\newblock Accessed: 2022-02-22.

\bibitem{bach2018design}
B.~Bach, Z.~Wang, M.~Farinella, D.~Murray-Rust, and N.~Henry~Riche.
\newblock Design patterns for data comics.
\newblock In {\em Proceedings of the CHI Conference on Human Factors in
  Computing Systems}, pp. 1--12. IEEE, New York, NY, 2018.
  \href{https://doi.org/10.1145/3173574.3173612}
{doi: {{%
10\hspace{.1pt}\discretionary{.}{%
}{.}\hspace{.4pt}1145\discretionary{/}{%
}{/}3173574\hspace{.1pt}\discretionary{.}{%
}{.}\hspace{.4pt}3173612}}}


\bibitem{baumgartl2021}
T.~Baumgartl, M.~Petzold, M.~Wunderlich, M.~Hohn, D.~Archambault, M.~Lieser,
  A.~Dalpke, S.~Scheithauer, M.~Marschollek, V.~M. Eichel, N.~T. Mutters,
  H.~Consortium, and T.~V. Landesberger.
\newblock In search of patient zero: Visual analytics of pathogen transmission
  pathways in hospitals.
\newblock {\em {IEEE} Transactions on Visualization and Computer Graphics},
  27(2):711--721, 2021. \href{https://doi.org/10.1109/TVCG.2020.3030437}
{doi: {{%
10\hspace{.1pt}\discretionary{.}{%
}{.}\hspace{.4pt}1109\discretionary{/}{%
}{/}TVCG\hspace{.1pt}\discretionary{.}{%
}{.}\hspace{.4pt}2020\hspace{.1pt}\discretionary{.}{%
}{.}\hspace{.4pt}3030437}}}


\bibitem{boy2015storytelling}
J.~Boy, F.~Detienne, and J.-D. Fekete.
\newblock Storytelling in information visualizations: Does it engage users to
  explore data?
\newblock In {\em Proceedings of the ACM Conference on Human Factors in
  Computing Systems}, pp. 1449--1458. ACM, New York, NY, 2015.
  \href{https://doi.org/10.1145/2702123}
{doi: {{%
10\hspace{.1pt}\discretionary{.}{%
}{.}\hspace{.4pt}1145\discretionary{/}{%
}{/}2702123}}}


\bibitem{boy2017showing}
J.~Boy, A.~V. Pandey, J.~Emerson, M.~Satterthwaite, O.~Nov, and E.~Bertini.
\newblock Showing people behind data: Does anthropomorphizing visualizations
  elicit more empathy for human rights data?
\newblock In {\em Proceedings of the CHI Conference on Human Factors in
  Computing Systems}, pp. 5462--5474. ACM, New York, NY, 2017.
  \href{https://doi.org/10.1145/3025453.3025512}
{doi: {{%
10\hspace{.1pt}\discretionary{.}{%
}{.}\hspace{.4pt}1145\discretionary{/}{%
}{/}3025453\hspace{.1pt}\discretionary{.}{%
}{.}\hspace{.4pt}3025512}}}


\bibitem{bradbury2020documentary}
J.~D. Bradbury and R.~E. Guadagno.
\newblock Documentary narrative visualization: Features and modes of
  documentary film in narrative visualization.
\newblock {\em Information Visualization}, 19(4):339--352, 2020.
  \href{https://doi.org/10.1177/1473871620925071}
{doi: {{%
10\hspace{.1pt}\discretionary{.}{%
}{.}\hspace{.4pt}1177\discretionary{/}{%
}{/}1473871620925071}}}


\bibitem{bran2010message}
R.~Bran.
\newblock Message in a bottle telling stories in a digital world.
\newblock {\em Procedia-Social and Behavioral Sciences}, 2(2):1790--1793, 2010.
  \href{https://doi.org/10.1016/j.sbspro.2010.03.986}
{doi: {{%
10\hspace{.1pt}\discretionary{.}{%
}{.}\hspace{.4pt}1016\discretionary{/}{%
}{/}j\hspace{.1pt}\discretionary{.}{%
}{.}\hspace{.4pt}sbspro\hspace{.1pt}\discretionary{.}{%
}{.}\hspace{.4pt}2010\hspace{.1pt}\discretionary{.}{%
}{.}\hspace{.4pt}03\hspace{.1pt}\discretionary{.}{%
}{.}\hspace{.4pt}986}}}


\bibitem{bryan2016temporal}
C.~Bryan, K.-L. Ma, and J.~Woodring.
\newblock Temporal summary images: An approach to narrative visualization via
  interactive annotation generation and placement.
\newblock {\em IEEE Transactions on Visualization and Computer Graphics},
  23(1):511--520, 2016. \href{https://doi.org/10.1109/TVCG.2016.2598876}
{doi: {{%
10\hspace{.1pt}\discretionary{.}{%
}{.}\hspace{.4pt}1109\discretionary{/}{%
}{/}TVCG\hspace{.1pt}\discretionary{.}{%
}{.}\hspace{.4pt}2016\hspace{.1pt}\discretionary{.}{%
}{.}\hspace{.4pt}2598876}}}


\bibitem{caiCharacterPlot}
Y.~Cai, C.~Miao, A.-H. Tan, and Z.~Shen.
\newblock A hybrid of plot-based and character-based interactive storytelling.
\newblock In {\em Technologies for E-Learning and Digital Entertainment}, pp.
  260--273. Springer Berlin Heidelberg, Berlin, Heidelberg, 2007.

\bibitem{cai2015applying}
Z.~Cai, Y.-N. Li, X.~S. Zheng, and K.~Zhang.
\newblock Applying feature integration theory to glyph-based information
  visualization.
\newblock In {\em IEEE Pacific Visualization Symposium}, pp. 99--103. IEEE, New
  York, NY, 2015. \href{https://doi.org/10.1109/PACIFICVIS.2015.7156363}
{doi: {{%
10\hspace{.1pt}\discretionary{.}{%
}{.}\hspace{.4pt}1109\discretionary{/}{%
}{/}PACIFICVIS\hspace{.1pt}\discretionary{.}{%
}{.}\hspace{.4pt}2015\hspace{.1pt}\discretionary{.}{%
}{.}\hspace{.4pt}7156363}}}


\bibitem{campbell2008hero}
J.~Campbell.
\newblock {\em The hero with a thousand faces}, vol.~17.
\newblock New World Library, 2008.

\bibitem{handwriting_2016}
S.~Carter, D.~Ha, I.~Johnson, and C.~Olah.
\newblock Experiments in handwriting with a neural network.
\newblock {\em Distill}, 2016. \href{https://doi.org/10.23915/distill.00004}
{doi: {{%
10\hspace{.1pt}\discretionary{.}{%
}{.}\hspace{.4pt}23915\discretionary{/}{%
}{/}distill\hspace{.1pt}\discretionary{.}{%
}{.}\hspace{.4pt}00004}}}


\bibitem{cavazza2001characters}
M.~Cavazza, F.~Charles, and S.~J. Mead.
\newblock Characters in search of an author: Ai-based virtual storytelling.
\newblock In {\em International Conference on Virtual Storytelling}, pp.
  145--154. Springer, 2001.

\bibitem{cavazza2001narrative}
M.~Cavazza, F.~Charles, and S.~J. Mead.
\newblock narrative representations and causality in character-based
  interactive storytelling.
\newblock {\em Proceedings of CAST 2001}, pp. 139--142, 2001.

\bibitem{charles2001character}
F.~Charles, S.~J. Mead, and M.~Cavazza.
\newblock Character-driven story generation in interactive storytelling.
\newblock In {\em Proceedings Seventh International Conference on Virtual
  Systems and Multimedia}, pp. 609--615. IEEE, 2001.

\bibitem{money_2014}
T.~Chu.
\newblock Money wins elections.
\newblock \small\url{http://letsfreecongress.org/}, 2014.

\bibitem{pixar22coats}
E.~Coats.
\newblock Pixar’s 22 rules of storytelling.
\newblock
  \small\url{https://www.aerogrammestudio.com/2013/03/07/pixars-22-rules-of-storytelling/}.
\newblock Accessed: 2022-02-22.

\bibitem{cruz2018process}
P.~Cruz, J.~Wihbey, A.~Ghael, F.~Shibuya, and S.~Costa.
\newblock Dendrochronology of u.s. immigration.
\newblock {\em Information Design Journal}, 25(1):6--20, 2019.
  \href{https://doi.org/10.1075/idj.25.1.01cru}
{doi: {{%
10\hspace{.1pt}\discretionary{.}{%
}{.}\hspace{.4pt}1075\discretionary{/}{%
}{/}idj\hspace{.1pt}\discretionary{.}{%
}{.}\hspace{.4pt}25\hspace{.1pt}\discretionary{.}{%
}{.}\hspace{.4pt}1\hspace{.1pt}\discretionary{.}{%
}{.}\hspace{.4pt}01cru}}}


\bibitem{dasu2018organic}
K.~Dasu, T.~Fujiwara, and K.-L. Ma.
\newblock An organic visual metaphor for public understanding of conditional
  co-occurrences.
\newblock In {\em 2018 IEEE Scientific Visualization Conference (SciVis)}, pp.
  1--5. IEEE, 2018.

\bibitem{dasu2020sea}
K.~Dasu, K.-L. Ma, J.~Ma, and J.~Frazier.
\newblock Sea of genes: A reflection on visualising metagenomic data for
  museums.
\newblock {\em IEEE Transactions on Visualization and Computer Graphics},
  27(2):935--945, 2020. \href{https://doi.org/10.1109/TVCG.2020.3030412}
{doi: {{%
10\hspace{.1pt}\discretionary{.}{%
}{.}\hspace{.4pt}1109\discretionary{/}{%
}{/}TVCG\hspace{.1pt}\discretionary{.}{%
}{.}\hspace{.4pt}2020\hspace{.1pt}\discretionary{.}{%
}{.}\hspace{.4pt}3030412}}}


\bibitem{infoaward}
{Data Visualization Society}.
\newblock {Information Is Beautiful Awards}.
\newblock {\small\url{https://www.informationisbeautifulawards.com/}}.
\newblock Accessed: 2023-06-29.

\bibitem{dudukovic2004telling}
N.~M. Dudukovic, E.~J. Marsh, and B.~Tversky.
\newblock Telling a story or telling it straight: The effects of entertaining
  versus accurate retellings on memory.
\newblock {\em Applied Cognitive Psychology}, 18(2):125--143, 2004.
  \href{https://doi.org/10.1002/acp.953}
{doi: {{%
10\hspace{.1pt}\discretionary{.}{%
}{.}\hspace{.4pt}1002\discretionary{/}{%
}{/}acp\hspace{.1pt}\discretionary{.}{%
}{.}\hspace{.4pt}953}}}


\bibitem{field2005screenplay}
S.~Field.
\newblock {\em Screenplay: The foundations of screenwriting}.
\newblock Delta, 2005.

\bibitem{fink2014dramatic}
E.~J. Fink.
\newblock {\em Dramatic story structure: A primer for screenwriters}.
\newblock Routledge, 2014.

\bibitem{forster2010aspects}
E.~M. Forster.
\newblock {\em Aspects of the Novel}.
\newblock RosettaBooks, 2010.

\bibitem{fritz2007hidden}
C.~Fritz and G.~Tosello.
\newblock The hidden meaning of forms: methods of recording paleolithic
  parietal art.
\newblock {\em Journal of archaeological method and theory}, 14(1):48--80,
  2007. \href{https://doi.org/10.1007/s10816-007-9027-3}
{doi: {{%
10\hspace{.1pt}\discretionary{.}{%
}{.}\hspace{.4pt}1007\discretionary{/}{%
}{/}s10816\discretionary{%
}{-}{-}007\discretionary{%
}{-}{-}9027\discretionary{%
}{-}{-}3}}}


\bibitem{gapminder}
GAPMINDER.ORG.
\newblock
  \small\url{https://www.gapminder.org/answers/will-saving-poor-children-lead-to-overpopulation/}.
\newblock Accessed: 2022-02-22.

\bibitem{gershon2001storytelling}
N.~Gershon and W.~Page.
\newblock What storytelling can do for information visualization.
\newblock {\em Communications of the ACM}, 44(8):31--37, 2001.
  \href{https://doi.org/10.1145/381641.381653}
{doi: {{%
10\hspace{.1pt}\discretionary{.}{%
}{.}\hspace{.4pt}1145\discretionary{/}{%
}{/}381641\hspace{.1pt}\discretionary{.}{%
}{.}\hspace{.4pt}381653}}}


\bibitem{gove2021automatic}
R.~Gove.
\newblock Automatic narrative summarization for visualizing cyber security logs
  and incident reports.
\newblock {\em {IEEE} Transactions on Visualization and Computer Graphics},
  28(1):1182--1190, 2021. \href{https://doi.org/10.1109/TVCG.2021.3114843}
{doi: {{%
10\hspace{.1pt}\discretionary{.}{%
}{.}\hspace{.4pt}1109\discretionary{/}{%
}{/}TVCG\hspace{.1pt}\discretionary{.}{%
}{.}\hspace{.4pt}2021\hspace{.1pt}\discretionary{.}{%
}{.}\hspace{.4pt}3114843}}}


\bibitem{howsure_2021}
N.~Halloran.
\newblock How sure are climate scientists, really?
\newblock \small\url{https://www.youtube.com/watch?v=R7FAAfK78_M}, 2021.

\bibitem{heer2007animated}
J.~Heer and G.~Robertson.
\newblock Animated transitions in statistical data graphics.
\newblock {\em IEEE Transactions on Visualization and Computer Graphics},
  13(6):1240--1247, 2007. \href{https://doi.org/10.1109/TVCG.2007.70539}
{doi: {{%
10\hspace{.1pt}\discretionary{.}{%
}{.}\hspace{.4pt}1109\discretionary{/}{%
}{/}TVCG\hspace{.1pt}\discretionary{.}{%
}{.}\hspace{.4pt}2007\hspace{.1pt}\discretionary{.}{%
}{.}\hspace{.4pt}70539}}}


\bibitem{hohman2020communicating}
F.~Hohman, M.~Conlen, J.~Heer, and D.~H.~P. Chau.
\newblock Communicating with interactive articles.
\newblock {\em Distill}, 5(9):e28, 2020.

\bibitem{hullman2011visualization}
J.~Hullman and N.~Diakopoulos.
\newblock Visualization rhetoric: Framing effects in narrative visualization.
\newblock {\em IEEE Transactions on Visualization and Computer Graphics},
  17(12):2231--2240, 2011. \href{https://doi.org/10.1109/TVCG.2011.255}
{doi: {{%
10\hspace{.1pt}\discretionary{.}{%
}{.}\hspace{.4pt}1109\discretionary{/}{%
}{/}TVCG\hspace{.1pt}\discretionary{.}{%
}{.}\hspace{.4pt}2011\hspace{.1pt}\discretionary{.}{%
}{.}\hspace{.4pt}255}}}


\bibitem{hullman2013contextifier}
J.~Hullman, N.~Diakopoulos, and E.~Adar.
\newblock Contextifier: automatic generation of annotated stock visualizations.
\newblock In {\em Proceedings of the SIGCHI Conference on Human Factors in
  Computing Systems}, pp. 2707--2716. ACM, New York, NY, 2013.
  \href{https://doi.org/10.1145/2470654.2481374}
{doi: {{%
10\hspace{.1pt}\discretionary{.}{%
}{.}\hspace{.4pt}1145\discretionary{/}{%
}{/}2470654\hspace{.1pt}\discretionary{.}{%
}{.}\hspace{.4pt}2481374}}}


\bibitem{hullman2013deeper}
J.~Hullman, S.~Drucker, N.~H. Riche, B.~Lee, D.~Fisher, and E.~Adar.
\newblock A deeper understanding of sequence in narrative visualization.
\newblock {\em IEEE transactions on visualization and computer graphics},
  19(12):2406--2415, 2013. \href{https://doi.org/10.1109/TVCG.2013.119}
{doi: {{%
10\hspace{.1pt}\discretionary{.}{%
}{.}\hspace{.4pt}1109\discretionary{/}{%
}{/}TVCG\hspace{.1pt}\discretionary{.}{%
}{.}\hspace{.4pt}2013\hspace{.1pt}\discretionary{.}{%
}{.}\hspace{.4pt}119}}}


\bibitem{outofsight2004}
P.~Interactive.
\newblock Out of sight, out of mind.
\newblock \small\url{https://drones.pitchinteractive.com/}.

\bibitem{isenberg2018immersive}
P.~Isenberg, B.~Lee, H.~Qu, and M.~Cordeil.
\newblock Immersive visual data stories.
\newblock In {\em Immersive Analytics}, pp. 165--184. Springer, New York, NY,
  2018.

\bibitem{kang2020role}
J.-A. Kang, S.~Hong, and G.~T. Hubbard.
\newblock The role of storytelling in advertising: Consumer emotion, narrative
  engagement level, and word-of-mouth intention.
\newblock {\em Journal of Consumer Behaviour}, 19(1):47--56, 2020.
  \href{https://doi.org/10.1002/cb.1793}
{doi: {{%
10\hspace{.1pt}\discretionary{.}{%
}{.}\hspace{.4pt}1002\discretionary{/}{%
}{/}cb\hspace{.1pt}\discretionary{.}{%
}{.}\hspace{.4pt}1793}}}


\bibitem{kim2017graphscape}
Y.~Kim, K.~Wongsuphasawat, J.~Hullman, and J.~Heer.
\newblock Graphscape: A model for automated reasoning about visualization
  similarity and sequencing.
\newblock In {\em Proceedings of the CHI Conference on Human Factors in
  Computing Systems}, pp. 2628--2638. ACM, New York, NY, 2017.
  \href{https://doi.org/10.1145/3025453.3025866}
{doi: {{%
10\hspace{.1pt}\discretionary{.}{%
}{.}\hspace{.4pt}1145\discretionary{/}{%
}{/}3025453\hspace{.1pt}\discretionary{.}{%
}{.}\hspace{.4pt}3025866}}}


\bibitem{NegativeEmo2022}
X.~Lan, Y.~Wu, Y.~Shi, Q.~Chen, and N.~Cao.
\newblock Negative emotions, positive outcomes? exploring the communication of
  negativity in serious data stories.
\newblock In {\em Proceedings of the 2022 CHI Conference on Human Factors in
  Computing Systems}. Association for Computing Machinery, New York, NY, USA,
  2022. \href{https://doi.org/10.1145/3491102.3517530}
{doi: {{%
10\hspace{.1pt}\discretionary{.}{%
}{.}\hspace{.4pt}1145\discretionary{/}{%
}{/}3491102\hspace{.1pt}\discretionary{.}{%
}{.}\hspace{.4pt}3517530}}}


\bibitem{lee2015more}
B.~Lee, N.~H. Riche, P.~Isenberg, and S.~Carpendale.
\newblock More than telling a story: Transforming data into visually shared
  stories.
\newblock {\em IEEE Computer Graphics and Applications}, 35(5):84--90, 2015.
  \href{https://doi.org/10.1109/MCG.2015.99}
{doi: {{%
10\hspace{.1pt}\discretionary{.}{%
}{.}\hspace{.4pt}1109\discretionary{/}{%
}{/}MCG\hspace{.1pt}\discretionary{.}{%
}{.}\hspace{.4pt}2015\hspace{.1pt}\discretionary{.}{%
}{.}\hspace{.4pt}99}}}


\bibitem{li2015metaphoric}
Y.-N. Li, D.-J. Li, and K.~Zhang.
\newblock Metaphoric transfer effect in information visualization using glyphs.
\newblock In {\em Proceedings of the International Symposium on Visual
  Information Communication and Interaction}, pp. 121--130. ACM, New York, NY,
  2015. \href{https://doi.org/10.1145/2801040.2801062}
{doi: {{%
10\hspace{.1pt}\discretionary{.}{%
}{.}\hspace{.4pt}1145\discretionary{/}{%
}{/}2801040\hspace{.1pt}\discretionary{.}{%
}{.}\hspace{.4pt}2801062}}}


\bibitem{liem2020structure}
J.~Liem, C.~Perin, and J.~Wood.
\newblock Structure and empathy in visual data storytelling: Evaluating their
  influence on attitude.
\newblock In {\em Computer Graphics Forum}, pp. 277--289. Wiley Online Library,
  2020.

\bibitem{ma2012scientific}
K.-L. Ma, I.~Liao, J.~Frazier, H.~Hauser, and H.-N. Kostis.
\newblock Scientific storytelling using visualization.
\newblock {\em IEEE Computer Graphics and Applications}, 32(1):12--19, 2012.
  \href{https://doi.org/10.1109/MCG.2012.24}
{doi: {{%
10\hspace{.1pt}\discretionary{.}{%
}{.}\hspace{.4pt}1109\discretionary{/}{%
}{/}MCG\hspace{.1pt}\discretionary{.}{%
}{.}\hspace{.4pt}2012\hspace{.1pt}\discretionary{.}{%
}{.}\hspace{.4pt}24}}}


\bibitem{VNF2017}
S.~McKenna, N.~Henry~Riche, B.~Lee, J.~Boy, and M.~Meyer.
\newblock Visual narrative flow: Exploring factors shaping data visualization
  story reading experiences.
\newblock {\em Computer Graphics Forum}, 36(3):377--387, 2017.
  \href{https://doi.org/10.1111/cgf.13195}
{doi: {{%
10\hspace{.1pt}\discretionary{.}{%
}{.}\hspace{.4pt}1111\discretionary{/}{%
}{/}cgf\hspace{.1pt}\discretionary{.}{%
}{.}\hspace{.4pt}13195}}}


\bibitem{morais2020showing}
L.~Morais, Y.~Jansen, N.~Andrade, and P.~Dragicevic.
\newblock Showing data about people: A design space of anthropographics.
\newblock {\em IEEE Transactions on Visualization and Computer Graphics},
  28(3):1661--1679, 2022. \href{https://doi.org/10.1109/TVCG.2020.3023013}
{doi: {{%
10\hspace{.1pt}\discretionary{.}{%
}{.}\hspace{.4pt}1109\discretionary{/}{%
}{/}TVCG\hspace{.1pt}\discretionary{.}{%
}{.}\hspace{.4pt}2020\hspace{.1pt}\discretionary{.}{%
}{.}\hspace{.4pt}3023013}}}


\bibitem{ojo2018patterns}
A.~Ojo and B.~Heravi.
\newblock Patterns in award winning data storytelling: Story types, enabling
  tools and competences.
\newblock {\em Digital journalism}, 6(6):693--718, 2018.
  \href{https://doi.org/10.1080/21670811.2017.1403291}
{doi: {{%
10\hspace{.1pt}\discretionary{.}{%
}{.}\hspace{.4pt}1080\discretionary{/}{%
}{/}21670811\hspace{.1pt}\discretionary{.}{%
}{.}\hspace{.4pt}2017\hspace{.1pt}\discretionary{.}{%
}{.}\hspace{.4pt}1403291}}}


\bibitem{petridis2019human}
S.~Petridis and L.~B. Chilton.
\newblock Human errors in interpreting visual metaphor.
\newblock In {\em Proceedings of on Creativity and Cognition}, p. 187–197.
  ACM, New York, NY, USA, 2019. \href{https://doi.org/10.1145/3325480.3325503}
{doi: {{%
10\hspace{.1pt}\discretionary{.}{%
}{.}\hspace{.4pt}1145\discretionary{/}{%
}{/}3325480\hspace{.1pt}\discretionary{.}{%
}{.}\hspace{.4pt}3325503}}}


\bibitem{ren2017chartaccent}
D.~Ren, M.~Brehmer, B.~Lee, T.~H{\"o}llerer, and E.~K. Choe.
\newblock Chartaccent: Annotation for data-driven storytelling.
\newblock In {\em IEEE Pacific Visualization Symposium}, pp. 230--239. IEEE,
  New York, NY, 2017. \href{https://doi.org/10.1109/PACIFICVIS.2017.8031599}
{doi: {{%
10\hspace{.1pt}\discretionary{.}{%
}{.}\hspace{.4pt}1109\discretionary{/}{%
}{/}PACIFICVIS\hspace{.1pt}\discretionary{.}{%
}{.}\hspace{.4pt}2017\hspace{.1pt}\discretionary{.}{%
}{.}\hspace{.4pt}8031599}}}


\bibitem{blast_2020}
Reuters.
\newblock How powerful was the beirut blast?
\newblock
  \small\url{https://graphics.reuters.com/LEBANON-SECURITY/BLAST/yzdpxnmqbpx/index.html},
  2020.

\bibitem{risch2008role}
J.~S. Risch.
\newblock On the role of metaphor in information visualization.
\newblock {\em arXiv preprint arXiv:0809.0884}, 2008.

\bibitem{sallaberry2016contact}
A.~Sallaberry, Y.-c. Fu, H.-C. Ho, and K.-L. Ma.
\newblock Contact trees: Network visualization beyond nodes and edges.
\newblock {\em PLOS ONE}, 11(1):1--23, 01 2016.
  \href{https://doi.org/10.1371/journal.pone.0146368}
{doi: {{%
10\hspace{.1pt}\discretionary{.}{%
}{.}\hspace{.4pt}1371\discretionary{/}{%
}{/}journal\hspace{.1pt}\discretionary{.}{%
}{.}\hspace{.4pt}pone\hspace{.1pt}\discretionary{.}{%
}{.}\hspace{.4pt}0146368}}}


\bibitem{sarica2016effect}
H.~{\c{C}}. Sar{\i}ca and Y.~K. Usluel.
\newblock The effect of digital storytelling on visual memory and writing
  skills.
\newblock {\em Computers \& Education}, 94:298--309, 2016.
  \href{https://doi.org/10.1016/j.compedu.2015.11.016}
{doi: {{%
10\hspace{.1pt}\discretionary{.}{%
}{.}\hspace{.4pt}1016\discretionary{/}{%
}{/}j\hspace{.1pt}\discretionary{.}{%
}{.}\hspace{.4pt}compedu\hspace{.1pt}\discretionary{.}{%
}{.}\hspace{.4pt}2015\hspace{.1pt}\discretionary{.}{%
}{.}\hspace{.4pt}11\hspace{.1pt}\discretionary{.}{%
}{.}\hspace{.4pt}016}}}


\bibitem{scarr_2020}
S.~Scarr, M.~Sharma, and G.~Bhatia.
\newblock {Covid-19}: The pace of death.
\newblock
  \small\url{https://www.reuters.com/graphics/HEALTH-CORONAVIRUS/DEATHS/xlbpgobgapq/index.html},
  Sep 2020.

\bibitem{segel2010narrative}
E.~Segel and J.~Heer.
\newblock Narrative visualization: Telling stories with data.
\newblock {\em IEEE Transactions on Visualization and Computer Graphics},
  16(6):1139--1148, 2010. \href{https://doi.org/10.1109/TVCG.2010.179}
{doi: {{%
10\hspace{.1pt}\discretionary{.}{%
}{.}\hspace{.4pt}1109\discretionary{/}{%
}{/}TVCG\hspace{.1pt}\discretionary{.}{%
}{.}\hspace{.4pt}2010\hspace{.1pt}\discretionary{.}{%
}{.}\hspace{.4pt}179}}}


\bibitem{shen2008edgar}
D.~Shen.
\newblock {Edgar Allan Poe's Aesthetic Theory, the Insanity Debate, and the
  Ethically Oriented Dynamics of ““The Tell-Tale Heart””}.
\newblock {\em Nineteenth-Century Literature}, 63(3):321--345, 12 2008.
  \href{https://doi.org/10.1525/ncl.2008.63.3.321}
{doi: {{%
10\hspace{.1pt}\discretionary{.}{%
}{.}\hspace{.4pt}1525\discretionary{/}{%
}{/}ncl\hspace{.1pt}\discretionary{.}{%
}{.}\hspace{.4pt}2008\hspace{.1pt}\discretionary{.}{%
}{.}\hspace{.4pt}63\hspace{.1pt}\discretionary{.}{%
}{.}\hspace{.4pt}3\hspace{.1pt}\discretionary{.}{%
}{.}\hspace{.4pt}321}}}


\bibitem{shi2020calliope}
D.~Shi, X.~Xu, F.~Sun, Y.~Shi, and N.~Cao.
\newblock Calliope: Automatic visual data story generation from a spreadsheet.
\newblock {\em {IEEE} Transactions on Visualization and Computer Graphics},
  27(2):453--463, 2020. \href{https://doi.org/10.1109/TVCG.2020.3030403}
{doi: {{%
10\hspace{.1pt}\discretionary{.}{%
}{.}\hspace{.4pt}1109\discretionary{/}{%
}{/}TVCG\hspace{.1pt}\discretionary{.}{%
}{.}\hspace{.4pt}2020\hspace{.1pt}\discretionary{.}{%
}{.}\hspace{.4pt}3030403}}}


\bibitem{shi2018meetingvis}
Y.~Shi, C.~Bryan, S.~Bhamidipati, Y.~Zhao, Y.~Zhang, and K.-L. Ma.
\newblock Meetingvis: Visual narratives to assist in recalling meeting context
  and content.
\newblock {\em IEEE Transactions on Visualization and Computer Graphics},
  24(6):1918--1929, 2018. \href{https://doi.org/10.1109/TVCG.2018.2816203}
{doi: {{%
10\hspace{.1pt}\discretionary{.}{%
}{.}\hspace{.4pt}1109\discretionary{/}{%
}{/}TVCG\hspace{.1pt}\discretionary{.}{%
}{.}\hspace{.4pt}2018\hspace{.1pt}\discretionary{.}{%
}{.}\hspace{.4pt}2816203}}}


\bibitem{srinivasan2018augmenting}
A.~Srinivasan, S.~M. Drucker, A.~Endert, and J.~Stasko.
\newblock Augmenting visualizations with interactive data facts to facilitate
  interpretation and communication.
\newblock {\em IEEE Transactions on Visualization and Computer Graphics},
  25(1):672--681, 2018. \href{https://doi.org/10.1109/TVCG.2018.2865145}
{doi: {{%
10\hspace{.1pt}\discretionary{.}{%
}{.}\hspace{.4pt}1109\discretionary{/}{%
}{/}TVCG\hspace{.1pt}\discretionary{.}{%
}{.}\hspace{.4pt}2018\hspace{.1pt}\discretionary{.}{%
}{.}\hspace{.4pt}2865145}}}


\bibitem{stolper2016emerging}
C.~D. Stolper, B.~Lee, N.~H. Riche, and J.~Stasko.
\newblock Emerging and recurring data-driven storytelling techniques: Analysis
  of a curated collection of recent stories.
\newblock Technical report, Microsoft Research, Washington, USA, 2016.

\bibitem{storyline2012}
Y.~Tanahashi and K.-L. Ma.
\newblock Design considerations for optimizing storyline visualizations.
\newblock {\em IEEE Transactions on Visualization and Computer Graphics},
  18(12):2679--2688, 2012. \href{https://doi.org/10.1109/TVCG.2012.212}
{doi: {{%
10\hspace{.1pt}\discretionary{.}{%
}{.}\hspace{.4pt}1109\discretionary{/}{%
}{/}TVCG\hspace{.1pt}\discretionary{.}{%
}{.}\hspace{.4pt}2012\hspace{.1pt}\discretionary{.}{%
}{.}\hspace{.4pt}212}}}


\bibitem{soccer_2014}
T.~N.~Y. Times.
\newblock The world's ball.
\newblock
  \small\url{https://www.nytimes.com/interactive/2014/06/13/sports/worldcup/world-cup-balls.html},
  2014.

\bibitem{tong2018storytelling}
C.~Tong, R.~Roberts, R.~Borgo, S.~Walton, R.~S. Laramee, K.~Wegba, A.~Lu,
  Y.~Wang, H.~Qu, Q.~Luo, and X.~Ma.
\newblock Storytelling and visualization: An extended survey.
\newblock {\em Information}, 9(3), 2018.
  \href{https://doi.org/10.3390/info9030065}
{doi: {{%
10\hspace{.1pt}\discretionary{.}{%
}{.}\hspace{.4pt}3390\discretionary{/}{%
}{/}info9030065}}}


\bibitem{truby2008anatomy}
J.~Truby.
\newblock {\em The anatomy of story: 22 steps to becoming a master
  storyteller}.
\newblock Farrar, Straus and Giroux, 2008.

\bibitem{wangNarvis2019}
Q.~Wang, Z.~Li, S.~Fu, W.~Cui, and H.~Qu.
\newblock Narvis: Authoring narrative slideshows for introducing data
  visualization designs.
\newblock {\em {IEEE} Transactions on Visualization and Computer Graphics},
  25(1):779--788, 2019. \href{https://doi.org/10.1109/TVCG.2018.2865232}
{doi: {{%
10\hspace{.1pt}\discretionary{.}{%
}{.}\hspace{.4pt}1109\discretionary{/}{%
}{/}TVCG\hspace{.1pt}\discretionary{.}{%
}{.}\hspace{.4pt}2018\hspace{.1pt}\discretionary{.}{%
}{.}\hspace{.4pt}2865232}}}


\bibitem{wang2015design}
S.~Wang, Y.~Tanahashi, N.~Leaf, and K.-L. Ma.
\newblock Design and effects of personal visualizations.
\newblock {\em IEEE Computer Graphics and Applications}, 35(4):82--93, 2015.
  \href{https://doi.org/10.1109/MCG.2015.74}
{doi: {{%
10\hspace{.1pt}\discretionary{.}{%
}{.}\hspace{.4pt}1109\discretionary{/}{%
}{/}MCG\hspace{.1pt}\discretionary{.}{%
}{.}\hspace{.4pt}2015\hspace{.1pt}\discretionary{.}{%
}{.}\hspace{.4pt}74}}}


\bibitem{wangDataShot2020}
Y.~Wang, Z.~Sun, H.~Zhang, W.~Cui, K.~Xu, X.~Ma, and D.~Zhang.
\newblock Datashot: Automatic generation of fact sheets from tabular data.
\newblock {\em {IEEE} Transactions on Visualization and Computer Graphics},
  26(1):895--905, 2020. \href{https://doi.org/10.1109/TVCG.2019.2934398}
{doi: {{%
10\hspace{.1pt}\discretionary{.}{%
}{.}\hspace{.4pt}1109\discretionary{/}{%
}{/}TVCG\hspace{.1pt}\discretionary{.}{%
}{.}\hspace{.4pt}2019\hspace{.1pt}\discretionary{.}{%
}{.}\hspace{.4pt}2934398}}}


\bibitem{WillikinWoolf}
W.~Woolf.
\newblock
  \small\url{https://twitter.com/WillikinWolf/status/1176006515968241665}.
\newblock Accessed: 2022-02-22.

\bibitem{yang2022}
L.~Yang, X.~Xu, X.~Lan, Z.~Liu, S.~Guo, Y.~Shi, H.~Qu, and N.~Cao.
\newblock A design space for applying the freytag's pyramid structure to data
  stories.
\newblock {\em IEEE Transactions on Visualization and Computer Graphics},
  28(1):922--932, 2022. \href{https://doi.org/10.1109/TVCG.2021.3114774}
{doi: {{%
10\hspace{.1pt}\discretionary{.}{%
}{.}\hspace{.4pt}1109\discretionary{/}{%
}{/}TVCG\hspace{.1pt}\discretionary{.}{%
}{.}\hspace{.4pt}2021\hspace{.1pt}\discretionary{.}{%
}{.}\hspace{.4pt}3114774}}}


\bibitem{9552203}
L.~Yang, X.~Xu, X.~Lan, Z.~Liu, S.~Guo, Y.~Shi, H.~Qu, and N.~Cao.
\newblock A design space for applying the freytag's pyramid structure to data
  stories.
\newblock {\em IEEE Transactions on Visualization and Computer Graphics},
  28(1):922--932, 2022. \href{https://doi.org/10.1109/TVCG.2021.3114774}
{doi: {{%
10\hspace{.1pt}\discretionary{.}{%
}{.}\hspace{.4pt}1109\discretionary{/}{%
}{/}TVCG\hspace{.1pt}\discretionary{.}{%
}{.}\hspace{.4pt}2021\hspace{.1pt}\discretionary{.}{%
}{.}\hspace{.4pt}3114774}}}


\bibitem{zhao2015data}
Z.~Zhao, R.~Marr, and N.~Elmqvist.
\newblock Data comics: Sequential art for data-driven storytelling.
\newblock Technical report, Univ. of Maryland, 2015.

\end{thebibliography}

\appendix   

\section{About Appendices}
Refer to \cref{sec:appendices_inst} for instructions regarding appendices.

\section{Troubleshooting}
\label{appendix:troubleshooting}

\subsection{ifpdf error}

If you receive compilation errors along the lines of \texttt{Package ifpdf Error: Name clash, \textbackslash ifpdf is already defined} then please add a new line \verb|\let\ifpdf\relax| right after the \verb|\documentclass[journal]{vgtc}| call.
Note that your error is due to packages you use that define \verb|\ifpdf| which is obsolete (the result is that \verb|\ifpdf| is defined twice); these packages should be changed to use \verb|ifpdf| package instead.

\subsection{\texttt{pdfendlink} error}

Occasionally (for some \LaTeX\ distributions) this hyper-linked bib\TeX\ style may lead to \textbf{compilation errors} (\texttt{pdfendlink ended up in different nesting level ...}) if a reference entry is broken across two pages (due to a bug in \verb|hyperref|).
In this case, make sure you have the latest version of the \verb|hyperref| package (i.e.\ update your \LaTeX\ installation/packages) or, alternatively, revert back to \verb|\bibliographystyle{abbrv-doi}| (at the expense of removing hyperlinks from the bibliography) and try \verb|\bibliographystyle{abbrv-doi-hyperref}| again after some more editing.

\end{document}